\documentclass[english]{scrartcl}
\usepackage[T1]{fontenc}
\usepackage[latin9]{inputenc}
\usepackage{geometry}
\usepackage{babel}
\usepackage{array}
\usepackage{rotfloat}
\usepackage{multirow}
\usepackage{amsmath}
\usepackage{graphicx}
\usepackage[authoryear]{natbib}
\usepackage[unicode=true,
 bookmarks=true,bookmarksnumbered=false,bookmarksopen=false,
 breaklinks=true,pdfborder={0 0 0},pdfborderstyle={},backref=false,colorlinks=false]
 {hyperref}
\hypersetup{pdftitle={Plant coexistence mediated by adaptive foraging preferences of exploiters  or mutualists},
 pdfborderstyle={},pdfborderstyle={},pdfborderstyle={},pdfborderstyle={},pdfborderstyle={},pdfborderstyle={},pdfborderstyle={},pdfborderstyle={}}
\usepackage{breakurl}

\makeatletter

\providecommand{\tabularnewline}{\\}

\usepackage{babel}
\usepackage{lineno}
\usepackage[labelfont=bf,font=footnotesize,labelsep=period,format=plain]{caption}
\usepackage[auth-lg]{authblk}

\author[1,2]{Vlastimil K\v{r}ivan\footnote{email: vlastimil.krivan@gmail.com}}

\author[2,1]{Tom\'{a}s A. Revilla\footnote{corresponding author, email: tomrevilla@gmail.com}}

\affil[1]{Department of Mathematics, Faculty of Science, University of South Bohemia, 
                Brani\v{s}ovsk\'{a} 1760, 370 05 \v{C}esk\'{e} Bud\v{e}jovice, Czech Republic}

\affil[2]{Czech Academy of Sciences, Biology Centre, 
Institute of Entomology,    Brani\v{s}ovsk\'{a} 31, 370 05 \v{C}esk\'{e} Bud\v{e}jovice, Czech Republic}

\makeatother

\begin{document}

\title{Plant coexistence mediated by adaptive foraging preferences of exploiters or mutualists}
\maketitle
\begin{abstract}
Coexistence of plants depends on their competition for common resources and indirect
interactions mediated by shared exploiters or mutualists. These interactions are
driven either by changes in animal abundance (density-mediated interactions, e.g.,
apparent competition), or by changes in animal preferences for plants (behaviorally-mediated
interactions). This article studies effects of behaviorally-mediated interactions
on two plant population dynamics and animal preference dynamics when animal densities
are fixed. Animals can be either adaptive exploiters or adaptive mutualists (e.g.,
herbivores or pollinators) that maximize their fitness. Analysis of the model shows
that adaptive animal preferences for plants can lead to multiple outcomes of plant
coexistence with different levels of specialization or generalism for the mediator
animal species. In particular, exploiter generalism promotes plant coexistence even
when inter-specific competition is too strong to make plant coexistence possible
without exploiters, and mutualist specialization promotes plant coexistence at alternative
stable states when plant inter-specific competition is weak. Introducing a new concept
of generalized isoclines allows us to fully analyze the model with respect to the
strength of competitive interactions between plants (weak or strong), and the type
of interaction between plants and animals (exploitation or mutualism).

~


~ 
\end{abstract}
Highlights: 
\begin{itemize}
\item Adaptive exploiters make coexistence of two strongly competing plant species possible. 
\item Adaptive mutualists promote alternative plant coexistence states under weak competition. 
\item Adaptive mutualists always specialize on a single plant. 
\item The theory extends the isocline concept for ecological models with adaptive traits. 
\end{itemize}

\section{Introduction}

How do competing species coexist has been a puzzling question for ecologists. The
competitive exclusion principle states that two species competing for the same resource
cannot coexist at an equilibrium \citep{gause1934,hardin-science60}. This view is
supported by the Lotka\textendash Volterra competition model which predicts that
coexistence requires inter-specific competition to be weaker than intra-specific
competition. The ecological interpretation is that niche overlap for competing species
cannot be too large for species coexistence at an equilibrium \citep{macarthur_levins-amnat67}.
These early models of competition focused on two species competing either directly,
or indirectly (i.e., interference vs. exploitative competition). Exploitative competition
is an example of indirect interaction between two populations mediated by common
resources \citep{grover1997}. Another indirect interaction is apparent competition
\citep{holt-tpb77} that is mediated by shared consumers. In these competitive scenarios
coexistence requires that species are limited by different factors. Thus, two exponentially
growing plants will not coexist if they are limited by the same resource (``$R^{*}$''
rule, \citealt{tilman1982}) or by the same single predator (``$P^{*}$'' rule,
\citealt{holt_etal-amnat94}). Plant\textendash animal mutualisms, on the other hand,
can lead to apparent facilitation as in the case of pollination \citep{feinsinger-tree87,ghazoul-joe06}
where two plants flowering in different times can sustain large pollinator populations
\citep{waser_real-nature79}.

Indirect interactions can be either density- or behaviorally-mediated. In density-mediated
indirect interactions the mediator species density changes. E.g., in apparent competition
an increase in one plant density increases herbivore density which, in turn, decreases
density of the other plant species. In behaviorally-mediated indirect interactions
changes in one plant population density are transmitted through changes in animal
behavior when animal population density is fixed. In reality, both density- and trait-mediated
indirect interactions operate concurrently \citep{bolker_etal-ecology03,krivan_schmitz-oikos04}.
Analysis of the apparent competition food web module with two plants and their common
consumers who undergo population dynamics and adaptively change their foraging preferences
showed that combination of density- and behaviorally-mediated interactions promotes
plant coexistence that would not be possible if consumer preferences were fixed \citep{krivan-amnat97}.
Even when consumers were kept at fixed densities but they adaptively changed their
preferences for plants, plant coexistence was still promoted by behaviorally-mediated
interactions only \citep{krivan-tpb03}. This suggests that in antagonistic networks
adaptive foraging promotes species coexistence by reducing apparent competition.
This was verified in more complex antagonistic di- and tri- trophic food web modules
with many species \citep{krivan-jtb10}. In simulated complex antagonistic food-webs
adaptive prey switching also led to increased species persistence \citep{kondoh-science03,berec_etal-jtb10}.

Antagonistic interactions such as competition, predation and parasitism are cornerstones
of the niche centric view of community structure (e.g., food webs, guilds), and theories
of ecological dynamics and biodiversity (e.g., stability\textendash complexity debate).
Currently, there is a great interest about the role of mutualisms as factors shaping
communities \citep{bastolla_etal-nature09,bronstein2015}. As it turns out, many
mutualisms are mediated by consumer\textendash resource mechanisms, and several of
them evolved from exploitative relationships such as parasitisms \citep{bronstein2015}.
Thus, we may be able to understand consequences of both mutualisms and antagonisms
using common methodologies \citep{holland_deangelis-ecology10}. Several models considered
apparent competition or apparent facilitation separately, and more recently, also
together in the context of mixed mutualistic\textendash antagonistic communities
\citep{mougi_kondoh-ecores14,sauve_etal-theorecol15}. A limited number of models
consider density- and behaviorally-mediated effects transmitted by mutualisms. Some
predict that adaptive mutualism promotes coexistence in the case of large communities
\citep{valdovinos_etal-oikos13,mougi_kondoh-ecores14}, while others predict that
adaptation constrain coexistence by favoring profitable partner species in detriment
to rare ones \citep{revilla_krivan-plosone16}. Thus, more research is required to
evaluate the importance of adaptation and plasticity as drivers of population dynamics
and community structure in interaction networks that combine both mutualistic and
antagonistic interactions. And this motivates us to explore how adaptive behavior
of exploiters or mutualists changes the outcomes of competition between the plants
with which they interact.

In this article we analyze how behaviorally-mediated interactions transmitted by
shared animals influence plant competition. We demonstrate that foraging behavior
of animal exploiters (e.g., herbivores) or mutualists (e.g., pollinators) can have
important and predictable consequences for plant competitive coexistence. By assuming
that animal population densities are fixed, we eliminate density-mediated effects,
e.g., apparent competition or apparent facilitation. In this way, we can focus entirely
on indirect effects that are mediated only by changes in animal preferences (i.e.,
they are trait-mediated) for plants. We give conditions for plant coexistence at
an equilibrium under exploitation or mutualism either when interaction strength is
fixed, or when it is adaptive and maximizes animal fitness.

A plant competition model with adaptive preferences of one animal species for two
plants is presented in Section 2. Because optimal animal strategy is not uniquely
defined when both plants provide the same payoffs to animals, plant population dynamics
are described by a differential inclusion \citep{aubincellina1984,colombo_krivan-jmamb93}.
For such models we introduce generalized isoclines that allow us to fully analyze
the model. Section 3 provides a complete classification of plant equilibria and corresponding
animal preferences when animals are either exploiters or mutualists and when inter-specific
plant competition is either weak or strong. We conclude that adaptive exploitation
permits global stable coexistence when competition between plants is weak, and global
or local stable coexistence when competition is strong. In the case of adaptive mutualism
only weakly competing plants can coexist at a single equilibrium or at one of two
alternative stable states.

\section{Model}

We consider an interaction module consisting of two competing plant species with
population densities $P_{1}$ and $P_{2}$ and one animal species with population
density $A$. The important feature of this interaction module is that plant\textendash animal
interactions can be either exploitative (e.g., folivory, granivory, modeled by parameter
$s=-1$) or mutualistic (e.g., pollination, seed dispersal, $s=1$). We assume that
animal population density $A$ is fixed, and we are interested in plant population
dynamics that are described by a Lotka\textendash Volterra (LV) model 
\begin{equation}
\begin{aligned}\frac{dP_{1}}{dt}=r_{1}\left(1-\frac{P_{1}+c_{2}P_{2}}{K_{1}}\right)P_{1}+su_{1}P_{1}A\\
\frac{dP_{2}}{dt}=r_{2}\left(1-\frac{P_{2}+c_{1}P_{1}}{K_{2}}\right)P_{2}+su_{2}P_{2}A
\end{aligned}
\label{eq:plant_odes}
\end{equation}
where $r_{i}>0$ and $K_{i}>0$ are plant intrinsic growth rates and environmental
carrying capacities in absence of inter-specific interactions, and $c_{i}\geq0$
is the competition coefficient that measures competitive effects of plant $i$ on
the other plant. The strength of plant\textendash animal interactions depends on
animal density $(A)$ as well as on animal preferences $u_{1}$ and $u_{2}$ for
plant 1 and 2, respectively ($u_{i}\geq0$ for $i=1,2$ and $u_{1}+u_{2}=1$). Preference
for plant $i$ can be interpreted as the proportion of time that an animal spends
interacting with that plant, or, alternatively, as the fraction of the animal population
$(u_{i}A)$ interacting with that plant.

When animals are mutualists ($s=1$), model (\ref{eq:plant_odes}) assumes facultative
mutualism for plants, i.e., plant populations can grow even without animals. This
is a reasonable assumption because the great majority of plants do not rely on a
single mutualist species. E.g., when the mutualist is a pollinator, plants can be
pollinated by other means (e.g., by wind, or another pollinator species that is not
being explicitly considered). Another feature of model (\ref{eq:plant_odes}) is
that it assumes constant animal density. This can be a reasonable assumption if plant
population dynamics are faster than animal population dynamics or model (\ref{eq:plant_odes})
describes plant dynamics in a small locality, saturated at level $A$ by a large
regional population of highly mobile animals \citep{melian_etal-oikos09}. In these
scenarios effects of plants on animal population density (i.e., the numeric response)
can be ignored. However, feedbacks between plant density and animal foraging behavior
can remain important. Animal adaptation in response to changes in plant community
composition affects animal fitness even when the numerical response is not considered.
In turn, changes in animal preference influence population density of plants and
alter plant community composition. The constant animal density assumption allows
us to focus on behavior-mediated effects arising from adaptive animal preferences
for plants.

For fixed animal preferences $u_{i}$ ($i=1,2$) model (\ref{eq:plant_odes}) is
the classical Lotka\textendash Volterra competitive system with well known dynamics
\citep[e.g.,][]{case2000,rohr_etal-science14}. In particular, both plants coexist
at a globally stable equilibrium

\begin{equation}
(\hat{P}_{1},\hat{P}_{2})=\left(\frac{K_{1}r_{2}(r_{1}+sAu_{1})-c_{2}K_{2}r_{1}(r_{2}+sAu_{2})}{r_{1}r_{2}(1-c_{1}c_{2})},\frac{K_{2}r_{1}(r_{2}+sAu_{2})-c_{1}K_{1}r_{2}(r_{1}+sAu_{1})}{r_{1}r_{2}(1-c_{1}c_{2})}\right)\label{eq:lvequilibrium}
\end{equation}
if and only if the ratio of carrying capacities satisfies\footnote{When $A=0$ inequalities (\ref{eq:K1K2_condition}) reduce to $c_{2}<\frac{K_{1}}{K_{2}}<\frac{1}{c_{1}}$
which are the classic conditions for stable coexistence in the Lotka\textendash Volterra
competition model. 
} 
\begin{equation}
\frac{c_{2}(1+su_{2}A/r_{2})}{1+su_{1}A/r_{1}}<\frac{K_{1}}{K_{2}}<\frac{1+su_{2}A/r_{2}}{c_{1}(1+su_{1}A/r_{1})}.\label{eq:K1K2_condition}
\end{equation}
Thus, stable plant coexistence requires that 
\begin{equation}
c_{1}c_{2}<1.\label{eq:c1c2_condition}
\end{equation}

When inequalities in (\ref{eq:K1K2_condition}) are reversed, equilibrium (\ref{eq:lvequilibrium})
is still feasible for intermediate $K_{1}/K_{2}$ ratios, but it is unstable, i.e.,
either plant 1 or 2 wins depending on initial conditions. This is the bi-stable outcome
for the LV model when inter-specific competition is stronger relative to intra-specific
competition $(c_{1}c_{2}>1)$. If under exploitation $u_{i}A>r_{i}$, plant $i$
is not viable and no interior equilibrium exists.

In the next sections we show that these predictions change when animals behave adaptively
and they maximize their fitness.

\subsection{Adaptive animal preferences}

Here we assume that animal preferences change in the direction that maximizes animal
fitness. The payoff to an animal when feeding on plant $i(=1,2)$ is measured, e.g.,
as the amount of energy obtained per unit of time, i.e., $e_{i}P_{i}$ where $e_{i}$
denotes the amount of energy obtained from a single plant per unit of time. Animal
fitness is then defined as the average payoff, i.e., $W_{A}=e_{1}u_{1}P_{1}+e_{2}u_{2}P_{2}$
where $u_{1}+u_{2}=1$ and $u_{i}\geq0$. Under the ideal circumstances where individuals
have a perfect knowledge about plant profitabilities and abundances, maximization
of this fitness leads to the following optimal foraging strategy \citep{krivan-tpb03,krivan_vrkoc-jmb07}:
\begin{equation}
u_{1}\in U_{1}(P_{1},P_{2})=\begin{cases}
\{0\} & \text{when}\quad e_{1}P_{1}<e_{2}P_{2}\\{}
[0,1] & \text{when}\quad e_{1}P_{1}=e_{2}P_{2}\\
\{1\} & \text{when}\quad e_{1}P_{1}>e_{2}P_{2}.
\end{cases}\label{eq:pref_step}
\end{equation}
When plant densities are such that 
\begin{equation}
e_{1}P_{1}=e_{2}P_{2},\label{eq:A-switch}
\end{equation}
animal preference for plant 1 ($u_{1}$) is not uniquely defined and can take any
value between 0 and 1. This is because either of the two plants provides the same
payoff for animals.

\begin{figure}
\begin{centering}
\includegraphics{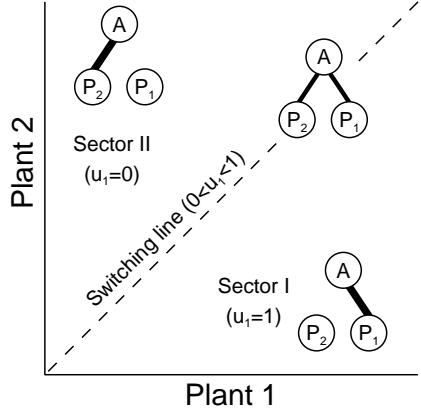} 
\par\end{centering}
\caption{\label{fig:modules}Interactions as a function of plant densities (axes) when animal
preference changes according to the step-like rule (\ref{eq:pref_step}). Below the
switching line (\ref{eq:A-switch}) animals specialize on plant 1, and above they
specialize on plant 2. Generalism occurs along the switching line where animals display
intermediate preferences for plants.}
\end{figure}

The switching line (\ref{eq:A-switch}) splits the positive quadrant of plant density
phase space in two sectors, as shown in Figure \ref{fig:modules}. In both of these
sectors, animals behave as specialists. In sector I (sector II), which is below (above)
the switching line, animals specialize on plant 1 (plant 2) only because this maximizes
their fitness. For plant densities along the switching line, animals have intermediate
preferences $(0<u_{1}<1)$, i.e., they are generalists that interact with both plants.

We observe that when $u_{1}$ is defined by (\ref{eq:pref_step}), model (\ref{eq:plant_odes})
becomes a differential inclusion, or, equivalently, a \citet{filippov1988} regularization
of a differential equation with a discontinuous right hand side (see Appendix \ref{sec:appdynsector};
\citealp{colombo_krivan-jmamb93}). To analyze such models we introduce in the next
section generalized isoclines.

\subsection{Interaction dynamics}

\subsubsection*{Generalized isoclines}

The effect of adaptive animals on plant coexistence can be predicted by isocline
analysis in the plant 1\textendash plant 2 phase plane. However, because population
dynamics (\ref{eq:plant_odes}) together with animal preferences (\ref{eq:pref_step})
are described by a differential inclusion, we need to define generalized plant isoclines
for this model. Isoclines need to be defined in both sectors I and II, as well as
in the switching line (\ref{eq:A-switch}).

Within sectors I or II plant 1 and 2 isoclines are 
\begin{equation}
\begin{aligned}P_{1}+c_{2}P_{2} & =H_{1}\\
P_{2}+c_{1}P_{1} & =H_{2},
\end{aligned}
\label{eq:isoclines}
\end{equation}
respectively. Here 
\begin{equation}
(H_{1}\,,\,H_{2})=\begin{cases}
\left(K_{1}(1+\frac{sA}{r_{1}}),\,K_{2}\right) & \textrm{in sector I}\\[0.2cm]
\left(K_{1},\,K_{2}(1+\frac{sA}{r_{2}})\right) & \textrm{in sector II}
\end{cases}\label{eq:Hi}
\end{equation}
are sector-dependent adjusted carrying capacities that depend on exploitative $(s=-1)$
or mutualistic animal effects $(s=1)$. For isoclines to exist in both sectors, $H_{1}$
and $H_{2}$ in (\ref{eq:Hi}) must be positive, i.e., $r_{i}+sA>0$, $i=1,2$. Plant
$i$ monoculture is viable under exploitation if $A<r_{i}$, i.e., plant $i$ has
limited tolerance for exploitation. If $A>r_{1}$ ($A>r_{2}$), isocline for plant
1 (plant 2) does not exist in sector I (sector II) under exploitation. On the other
hand, monocultures are always viable under facultative mutualism $(r_{i}+A>0)$.

\begin{figure}
\begin{centering}
\includegraphics[height=0.9\textheight]{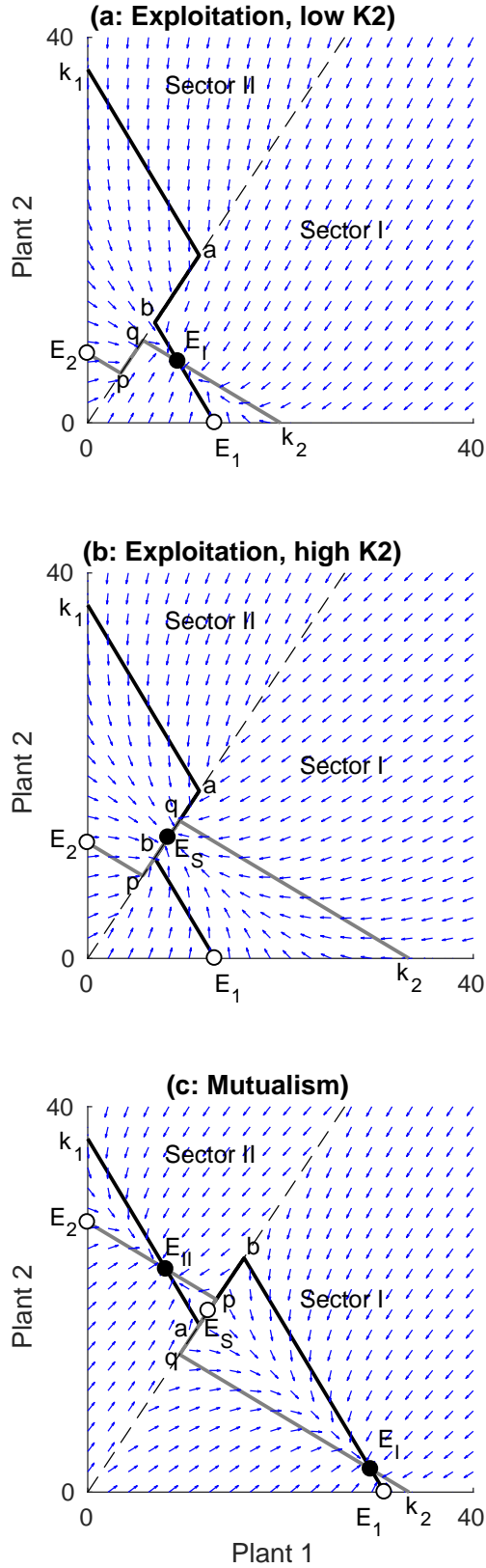} 
\par\end{centering}
\caption{\label{fig:GIclines}Generalized isoclines (plant 1: black, plant 2: gray) and plant
dynamics under weak competition ($c_{1}c_{2}<1)$. The (dashed) switching line (\ref{eq:A-switch})
splits the phase plane in sectors I and II. Stable equilibria are shown as dots,
and unstable equilibria as circles. Panel a assumes low population of exploiters
and isoclines intersect at a stable equilibrium in sector I. As the number of exploiters
increases (panel b), plants coexist at stable equilibrium $\mathbf{E_{S}}$ at the
switching line where animals are generalists. In panel c animals are mutualists and
isoclines intersect at two stable equilibria, one in each sector. Equilibrium $\mathbf{E_{S}}$
is unstable. Parameter values: $r_{i}=0.1,c_{i}=0.6,e_{1}=1.5,e_{2}=1,A=0.04,K_{1}=22$;
$K_{2}=12$ in (a) and $K_{2}=20$ in (b,c); $s=-1$ in panels a,b, and $s=1$ in
panel c.}
\end{figure}

As a result, isoclines in sectors I and II are piece-wise linear as illustrated in
Figure \ref{fig:GIclines}. Plant 1 isocline in sector I is the line segment connecting
points $\mathbf{b}$ and $\mathbf{E_{1}}$, and in sector II is the line segment
connecting points $\mathbf{k_{1}}$ and $\mathbf{a}$. Point 
\begin{equation}
\mathbf{E_{1}}=(P_{1}^{*},0)=\left(K_{1}\left(1+\frac{sA}{r_{1}}\right)\,,\,0\right)\label{eq:E1}
\end{equation}
is plant 1 monoculture equilibrium of model (\ref{eq:plant_odes}), and formulas
for intersection points \textbf{a}, \textbf{b} (with switching line) and $\mathbf{k_{1}}$
(with $P_{2}$ axis) are given in Appendix \ref{sec:appdynsector}. Similarly, plant
2 isocline consists of line segments connecting points $\mathbf{E_{2}}$ and \textbf{p}
in sector II, and \textbf{q} and $\mathbf{k_{2}}$ in sector I. Point 
\begin{equation}
\mathbf{E_{2}}=(0,P_{2}^{*})=\left(0\,,\,K_{2}\left(1+\frac{sA}{r_{2}}\right)\right)\label{eq:E2}
\end{equation}
is plant 2 monoculture equilibrium of model (\ref{eq:plant_odes}), and intersection
points \textbf{p}, \textbf{q} (with switching line) and $\mathbf{k_{2}}$ (with $P_{1}$
axis) are given in Appendix \ref{sec:appdynsector}. We remark that both monoculture
equilibria exist for mutualists while for exploiters, plant $i$ monoculture equilibrium
exists if $A<r_{i}.$

We define \emph{generalized isoclines} by adding the segment \textbf{a}\textemdash \textbf{b}
to plant 1 isocline, and segment \textbf{p}\textemdash \textbf{q} to plant 2 isocline.
Thus, both plant isoclines are continuous, piece-wise linear curves in plant phase
space. Plant 1 (plant 2) isocline is shown as the black (gray) line in Figure \ref{fig:GIclines}.
We stress here, that along their central segments (\textbf{a}\textemdash \textbf{b}
for plant 1 isocline, and \textbf{p}\textemdash \textbf{q} for plant 2 isocline)
the usual definition of isoclines as points of zero growth for particular plant species
does not hold for generalized isoclines. In particular, we show in the next section
that when the two segments partially overlap along the switching line as in Figure
\ref{fig:GIclines}b, c, the overlap segment (\textbf{b}\textemdash \textbf{q} in
panel b and \textbf{a}\textemdash \textbf{p} in panel c) does not consist of equilibria
only, as we explain in the next section.

We remark that under exploitation $(s=-1)$ plant 1 (plant 2) generalized isocline
consists of three segments if $r_{1}>A$ $(r_{2}>A)$. Otherwise, the isocline has
only two segments because $\mathbf{E_{1}}$ and \textbf{b} ($\mathbf{E_{2}}$ and
\textbf{p}) are not in the first quadrant. In case of mutualism $(s=1)$ generalized
isoclines always consist of three segments because monocultures are viable since
we assume that mutualism is facultative.

Appendix \ref{sec:appgradual} shows that generalized isoclines obtained for step-like
preferences given in (\ref{eq:pref_step}) are well approximated by smooth (usual)
isoclines when preferences are more gradual. However, the generalized isoclines allow
us to fully analyze the model.

\subsubsection*{Model equilibria}

In the classic Lotka\textendash Volterra (LV) model (\ref{eq:plant_odes}) stable
plant coexistence requires that the missing species can invade when the other plant
is at its population equilibrium. This is a consequence of linear isoclines that
generically intersect at most once. The case where animals behave adaptively is more
complex, because generalized isoclines are piece-wise linear and there can be interior
equilibria in both sectors (e.g., Figure \ref{fig:GIclines}c). In addition, we show
in this section that there is one equilibrium at the segment of the switching line
where the two isoclines coincide (e.g., Figure \ref{fig:GIclines}b, c).

We start by analyzing position of isoclines in sectors I and II. Since isoclines
are linear there they can intersect in either sector at most once. If they intersect,
the corresponding equilibrium is locally stable\footnote{By local stability we mean local asymptotic stability throughout this article.}
when $c_{1}c_{2}<1$ and unstable when $c_{1}c_{2}>1$. This follows from analysis
of the classic LV competition model. We also observe that at these equilibria animals
behave as specialists as they interact with a single plant only. To determine if
isoclines intersect within a sector, we compare their intersections with the corresponding
sector's axis and with the switching line (\ref{eq:A-switch}). In sector I we compare
position of equilibrium $\mathbf{E_{1}}$ with respect to point $\mathbf{k_{2}}$
on $P_{1}$ axis, and position of point \textbf{b} with respect to point \textbf{q}
on the switching line. If $\mathbf{E_{1}}$ exists and 
\begin{equation}
\mathbf{E_{1}}<\mathbf{k_{2}}\quad\text{and}\quad\textbf{q}<\textbf{b}\label{eq:con1}
\end{equation}
by which we mean that point $\mathbf{E_{1}}$ is to the left of point $\mathbf{\mathbf{k_{2}}}$
on $P_{1}$ axes and point \textbf{q} is to the left and down from point \textbf{b}
along the line $e_{1}P_{1}=e_{2}P_{2}$, or 
\begin{equation}
\mathbf{E_{1}}>\mathbf{k_{2}}\quad\text{and}\quad\textbf{q}>\textbf{b},\label{eq:con2}
\end{equation}

\noindent Appendix \ref{sec:appdynsector} shows that there is one coexistence equilibrium
\begin{equation}
\mathbf{E_{I}}=\left(\hat{P}_{1},\hat{P}_{2}\right)=\left(\frac{K_{1}r_{2}(r_{1}+sA)-c_{2}K_{2}r_{1}r_{2}}{r_{1}r_{2}(1-c_{1}c_{2})}\,,\,\frac{K_{2}r_{1}r_{2}-c_{1}K_{1}r_{2}(r_{1}+sA)}{r_{1}r_{2}(1-c_{1}c_{2})}\right)\label{eq:E.I}
\end{equation}
in sector I and this equilibrium is locally stable when (\ref{eq:con1}) holds because
in this case $c_{1}c_{2}<1$ (Figure \ref{fig:GIclines}a, c). If conditions in (\ref{eq:con2})
hold, the equilibrium is unstable. Appendix \ref{sec:appdynsector} shows that (\ref{eq:con1})
is equivalent with 
\begin{equation}
\gamma_{2}\equiv c_{1}\left(1+\frac{sA}{r_{1}}\right)<\frac{K_{2}}{K_{1}}<\left(\frac{e_{1}+c_{1}e_{2}}{e_{2}+c_{2}e_{1}}\right)\left(1+\frac{sA}{r_{1}}\right)\equiv\tau_{2}.\label{eq:coexSectorI}
\end{equation}
If both inequalities in (\ref{eq:coexSectorI}) are reversed, $\mathbf{E_{I}}$ still
exists because isoclines intersect in sector I but the equilibrium is unstable. If
$K_{2}/K_{1}$ is larger or smaller than both $\gamma_{2}$ and $\tau_{2}$, there
is no equilibrium in sector I because the two isoclines do not intersect there (e.g.,
Figure \ref{fig:GIclines}b where $\mathbf{E_{1}}<\mathbf{k_{2}}$ but \textbf{b}
< \textbf{q}).

Similarly, in sector II we compare position of $\mathbf{k_{1}}$ with respect to
equilibrium $\mathbf{E_{2}}$ on the $P_{2}$ axis, and position of \textbf{a} with
respect to \textbf{p} along the switching line. If equilibrium $\mathbf{E_{2}}$
exists and 
\begin{equation}
\mathbf{E_{2}<\mathbf{k_{1}}\quad\text{and}\quad\textbf{a}<\textbf{p}}\label{eq:con3}
\end{equation}
or 
\begin{equation}
\mathbf{E_{2}}>\mathbf{k_{1}}\quad\text{and}\quad\textbf{a}>\textbf{p},\label{eq:con4}
\end{equation}
Appendix \ref{sec:appdynsector} shows that there is one equilibrium in sector II
\begin{equation}
\mathbf{E_{II}}=\left(\hat{P}_{1},\hat{P}_{2}\right)=\left(\frac{K_{1}r_{1}r_{2}-c_{2}K_{2}r_{1}(r_{2}+sA)}{r_{1}r_{2}(1-c_{1}c_{2})}\,,\,\frac{K_{2}r_{1}(r_{2}+sA)-c_{1}K_{1}r_{1}r_{2}}{r_{1}r_{2}(1-c_{1}c_{2})}\right).\label{eq:E.II}
\end{equation}
This equilibrium is locally stable if and only if 
\begin{equation}
\gamma_{1}\equiv c_{2}\left(1+\frac{sA}{r_{2}}\right)<\frac{K_{1}}{K_{2}}<\left(\frac{e_{2}+c_{2}e_{1}}{e_{1}+c_{1}e_{2}}\right)\left(1+\frac{sA}{r_{2}}\right)\equiv\tau_{1}.\label{eq:coexSectorII}
\end{equation}
As in sector I, instability of $\mathbf{E_{II}}$ follows from inequality reversal
in (\ref{eq:coexSectorII}). When $K_{1}/K_{2}$ is larger or smaller than both $\gamma_{1}$
and $\tau_{1}$, no equilibrium exists in sector II.

We note that $\gamma_{1}$ and $\gamma_{2}$ are \emph{invasion thresholds} that
must be met by $K_{1}/K_{2}$ and $K_{2}/K_{1}$, respectively, for plant 1 to invade
at equilibrium $\mathbf{E_{2}}$ and for plant 2 to invade at equilibrium $\mathbf{E_{1}}$,
respectively. Invasion thresholds depend on resident plant parameters, plant\textendash animal
interaction type, and animal density. For example, $\gamma_{1}$ is directly proportional
to the competitive effect of plant 2 on plant 1 $(c_{2})$ exactly as in standard
LV models. This means that increasing inter-specific competition makes plant 1 less
likely to invade resident population consisting of plant 2 only. Under exploitation,
increasing animal density decreases the threshold allowing plant 1 to invade, while
increasing plant 2 intrinsic growth rate $(r_{2})$ makes this plant more difficult
to invade. These predictions change under mutualism because the invasion threshold
for plant 1 increases with increasing density of mutualists and decreases with plant
2 intrinsic growth rate.

Now we look for plant equilibria in the segment of the switching line where the two
generalized isoclines overlap. To answer this question we have to analyze plant dynamics
(\ref{eq:plant_odes}) with optimal animal behavior (\ref{eq:pref_step}) on the
switching line where animal preference for either plant is not uniquely defined.
Analysis in Appendix \ref{sec:appsliding} shows that when the two generalized isoclines
partially overlap along the switching line, there exists a single equilibrium in
the overlap segment 
\begin{equation}
\mathbf{E_{S}}=(\bar{P}_{1},\bar{P}_{2})=\left(\dfrac{e_{2}K_{1}K_{2}(r_{1}+r_{2}+sA)}{K_{1}r_{2}(e_{1}+c_{1}e_{2})+K_{2}r_{1}(e_{2}+c_{2}e_{1})}\,,\,\dfrac{e_{1}K_{1}K_{2}(r_{1}+r_{2}+sA)}{K_{1}r_{2}(e_{1}+c_{1}e_{2})+K_{2}r_{1}(e_{2}+c_{2}e_{1})}\right),\label{eq:E.S}
\end{equation}
see Figure \ref{fig:GIclines}b, c. This equilibrium is locally stable under exploitation
(Figure \ref{fig:GIclines}b) and unstable under mutualism (Figure \ref{fig:GIclines}c).
Appendix \ref{sec:appsliding} also shows that animal preference for plant 1 at this
equilibrium is 
\begin{equation}
\bar{u}_{1}=\dfrac{K_{2}r_{1}(r_{2}+sA)(e_{2}+c_{2}e_{1})-K_{1}r_{1}r_{2}(e_{1}+c_{1}e_{2})}{sA\left[K_{1}r_{2}(e_{1}+c_{1}e_{2})+K_{2}r_{1}(e_{2}+c_{2}e_{1})\right]},\label{eq:U.S}
\end{equation}
i.e., animals behave as generalists at this equilibrium.

\begin{figure}
\begin{centering}
\includegraphics[scale=0.7]{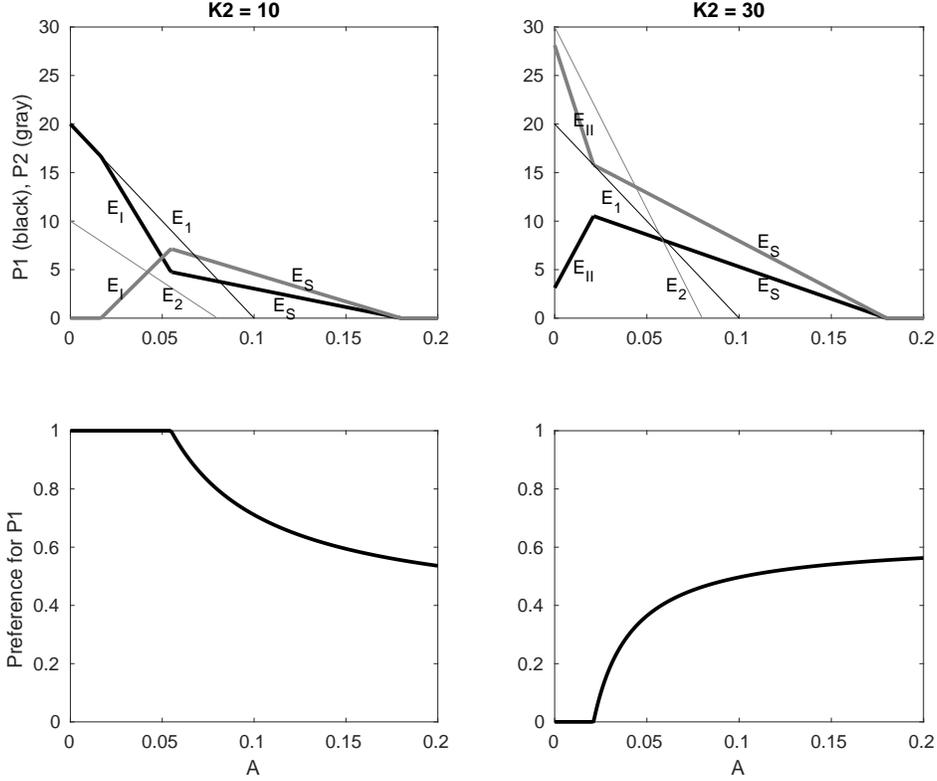} 
\par\end{centering}
\caption{\label{fig:coex_with_expl}Plant coexistence under weak competition $(c_{1}c_{2}<1)$
and adaptive exploitation $(s=-1)$, for high (left column) or low (right column)
$K_{1}/K_{2}$ ratios. Top panels show stable plant coexistence (thick lines) and
monoculture ($\mathbf{E_{1}}$ and $\mathbf{E_{2}}$, thin lines) equilibria as a
function of exploiter density. Bottom panels show corresponding exploiter preference
for plant 1 $(u_{1})$. It is interesting to observe that plant $i(=1,2)$ monoculture
is viable only when $A\leq r_{i}$ while when together, both plants form viable community
for higher animal densities satisfying $A\leq r_{1}+r_{2}$. Parameter values: $r_{1}=0.1,r_{2}=0.08,c_{i}=0.6,e_{1}=1.5,e_{2}=1,K_{1}=20$.}
\end{figure}

This analysis allows us to give meaning to \emph{attraction thresholds} $\tau_{i}$
defined in (\ref{eq:coexSectorI}) and (\ref{eq:coexSectorII}). For equilibrium
$\mathbf{E_{S}}$ to exist, $\bar{u}_{1}$ must be between 0 and 1. Under exploitation
($s=-1$) this happens when 
\begin{equation}
\frac{K_{1}}{K_{2}}>\tau_{1}\quad\textrm{and}\quad\frac{K_{2}}{K_{1}}>\tau_{2}\label{eq:cond_attract}
\end{equation}
while under mutualism ($s=1$) the conditions are 
\begin{equation}
\frac{K_{1}}{K_{2}}<\tau_{1}\quad\textrm{and}\quad\frac{K_{2}}{K_{1}}<\tau_{2}.\label{eq:cond_repel}
\end{equation}
Equilibrium $\mathbf{E_{S}}$ exists when $r_{1}+r_{2}+sA>0$ and it is always locally
stable for exploitation (i.e., (\ref{eq:cond_attract}) holds and ``$\mathbf{E_{S}}$
attracts'' locally trajectories from both sectors) and unstable for mutualism (i.e.,
(\ref{eq:cond_repel}) holds and ``$\mathbf{E_{S}}$ repels'' trajectories away;
see Appendix \ref{sec:appsliding}). If only one attraction threshold is passed,
equilibrium $\mathbf{E_{S}}$ does not exist and there is no plant population equilibrium
at which animals behave as generalists. Here the important observation is that existence
and stability of equilibrium $\mathbf{E_{S}}$ does not depend whether single plant
monocultures are viable or not. In fact, even if neither of the two plants is viable
(i.e., $A>r_{i}$, $i=1,2$), equilibrium $\mathbf{E_{S}}$ still exists provided
$A<r_{1}+r_{2}$ (Figure \ref{fig:coex_with_expl}). We show next how plant coexistence
and animal preferences depend on animal abundance and model parameters.

Figure \ref{fig:coex_with_expl} illustrates the effects of adaptive exploiters on
plant equilibria and exploiter preferences for plants. Let us consider the situation
where 
\begin{equation}
\frac{K_{1}}{K_{2}}>\frac{e_{2}+c_{2}e_{1}}{e_{1}+c_{1}e_{2}}\label{eqK1K2}
\end{equation}
(left column of Figure \ref{fig:coex_with_expl}). Without exploiters ($A=0$), plant
1 wins competition over plant 2. As the number of exploiters increases, exploiters
are plant 1 specialists ($u_{1}=1$, bottom-left panel) and plant 1 equilibrium density
decreases until $A\approx0.017$. For higher exploiter density (approx. $0.017<A<0.055$)
plant 2 invades plant 1 monoculture and both plants coexist at equilibrium $\mathbf{E_{I}}$
given in (\ref{eq:E.I}). Plant 1 population density keeps decreasing with increasing
$A$ while plant 2 population density increases. Exploiters still behave as specialists
on plant 1 till their population reaches another critical threshold $A\approx0.055$.
For yet higher exploiter density, animals behave as generalists feeding on both plants
with decreasing preference for plant 1 given in (\ref{eq:U.S}) and plants coexist
at equilibrium $\mathbf{E_{S}}$ given in (\ref{eq:E.S}). Thus, both plant population
densities now decrease with increasing animal abundance. The case where opposite
inequality in (\ref{eqK1K2}) holds is shown in the right panels of Figure \ref{fig:coex_with_expl}.
In this case, exploiters start as plant 2 specialists ($u_{1}=0$, bottom-right panel)
at plant equilibrium $\mathbf{E_{II}}$ given in (\ref{eq:E.II}). Thus, plant 2
decreases monotonically while plant 1 increases for $0\leq A<0.021$. Once both plants
are equally profitable for animals, animals become generalists and both plants start
to decrease together as preference for plant 1 keeps increasing.

Figure \ref{fig:coex_with_expl} also shows that adaptive exploitation leads to indirect
positive effects between plants. First, when animals are adaptive exploiters, plant
equilibrium densities are positive for animal densities at which plant monocultures
are not viable. E.g., plant 1 (plant 2) monoculture cannot exist for $A>0.1$ ($A>0.08$)
in Figure \ref{fig:coex_with_expl} but both plants do coexist at $\mathbf{E_{S}}$
as long as $A\leq r_{1}+r_{2}=0.18$. Thus, for large exploiter densities viability
of plant 1 relies on co-occurrence with plant 2 and vice-versa. Second, from (\ref{eq:E.S})
it follows that under generalism increasing $K_{1}$ or $K_{2}$ raises both plant
equilibrium densities (cf.~right vs.~left top panels in Figure \ref{fig:coex_with_expl}
for $A>0.05$). This is unlike standard LV models where increasing $K_{2}$ causes
increase of plant 2 equilibrium density and decrease of plant 1. The effect of other
parameters on plant equilibria $(\mathbf{E_{I}},\mathbf{E_{II}},\mathbf{E_{S}})$
is given in Appendix \ref{sec:appareffect}.

\begin{figure}
\begin{centering}
\includegraphics[scale=0.7]{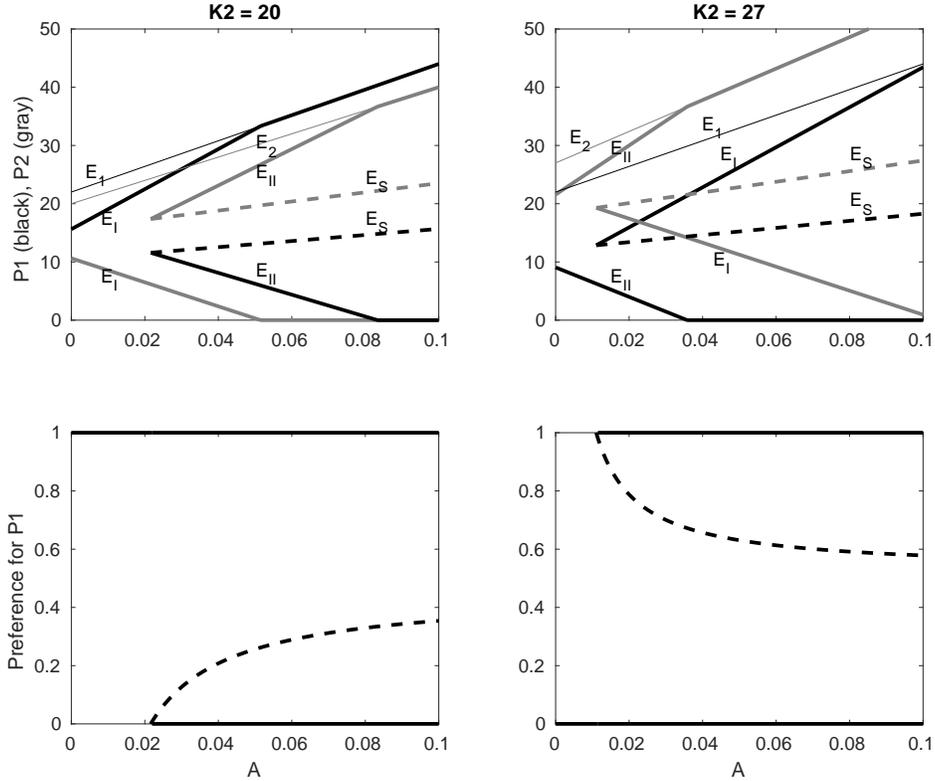} 
\par\end{centering}
\caption{\label{fig:coex_with_mut} Plant coexistence under weak competition $(c_{1}c_{2}<1)$
and adaptive mutualism $(s=1)$, for high (left column) or low (right column) $K_{1}/K_{2}$
ratios. Top panels show stable coexistence ($\mathbf{E_{I}}$ and $\mathbf{E_{II}}$,
thick lines) and monoculture ($\mathbf{E_{1}}$ and $\mathbf{E_{2}}$, thin lines)
equilibria as a function of exploiter density. Bottom panels show corresponding mutualist
preference for plant 1 $(u_{1})$. An alternative stable state (thick gray lines)
emerges when $u_{1}$ changes from 1 or 0 into $0<u_{1}<1$. Parameter values: $r_{i}=0.1,c_{i}=0.6,e_{1}=1.5,e_{2}=1,K_{1}=22$.}
\end{figure}

Effects of changes in parameters on plant equilibria in the case of mutualism are
often in opposite directions as compared to exploiters (see Appendix \ref{sec:appareffect}).
Because we assume that mutualism is facultative, plant monocultures ($\mathbf{E_{1}}$
and $\mathbf{E_{2}}$) are always viable and they increase with $A$. Provided both
plants coexist, plant 1 increases and plant 2 decreases with $A$ at equilibrium
$\mathbf{E_{I}}$, and the opposite happens at equilibrium $\mathbf{E_{II}}$. Equilibrium
$\mathbf{E_{S}}$, if it exists, is always unstable. 
Figure \ref{fig:coex_with_mut} serves as a good illustration. The left column displays
plant coexistence at equilibrium $\mathbf{E_{I}}$ when $A<0.022$ and animals specialize
on plant 1 ($u_{1}=1$). For higher animal densities there are two stable equilibria
$\mathbf{E_{I}}$ and $\mathbf{E_{II}}$ and unstable interior equilibrium $\mathbf{E_{S}}$
at which animals are generalists. The right column shows situation where $K_{1}/K_{2}$
is lower and plants coexists at equilibrium $\mathbf{E_{II}}$ when $A<0.0115$ and
animals specialize on plant 2. For higher animal densities there are two coexisting
stable plant equilibria $\mathbf{E_{I}}$ and $\mathbf{E_{II}}$ and the unstable
equilibrium $\mathbf{E_{S}}$.

\section{Plant coexistence under exploitation or mutualism}

By comparing $K_{1}/K_{2}$ with $\gamma_{1}$ and $\tau_{1}$ thresholds in (\ref{eq:coexSectorII}),
and $K_{2}/K_{1}$ with $\gamma_{2}$ and $\tau_{2}$ thresholds in (\ref{eq:coexSectorI}),
we provide a complete classification of model outcomes for all generic parameter
combinations, see Appendix \ref{sec:appclassification}. In the following sections
we discuss all possible global dynamics when animals are exploiters or mutualists,
and plant inter-specific competition is weak or strong. In the particular case of
exploitation, we only display scenarios where $A<r_{1}$ and $A<r_{2}$, i.e., plant
monocultures are viable and generalized isoclines display three segments. Scenarios
where monocultures are not viable, i.e., $A>r_{i}$, lead to similar global dynamics
as long as $r_{1}+r_{2}>A$ (i.e., if $A>r_{1}+r_{2}$ both plants go extinct like
in Figure \ref{fig:coex_with_expl}).

\subsection{Exploitation ($s=-1$) and weak inter-specific plant competition ($c_{1}c_{2}<1$)}

\begin{figure}
\begin{centering}
\includegraphics[width=1\textwidth]{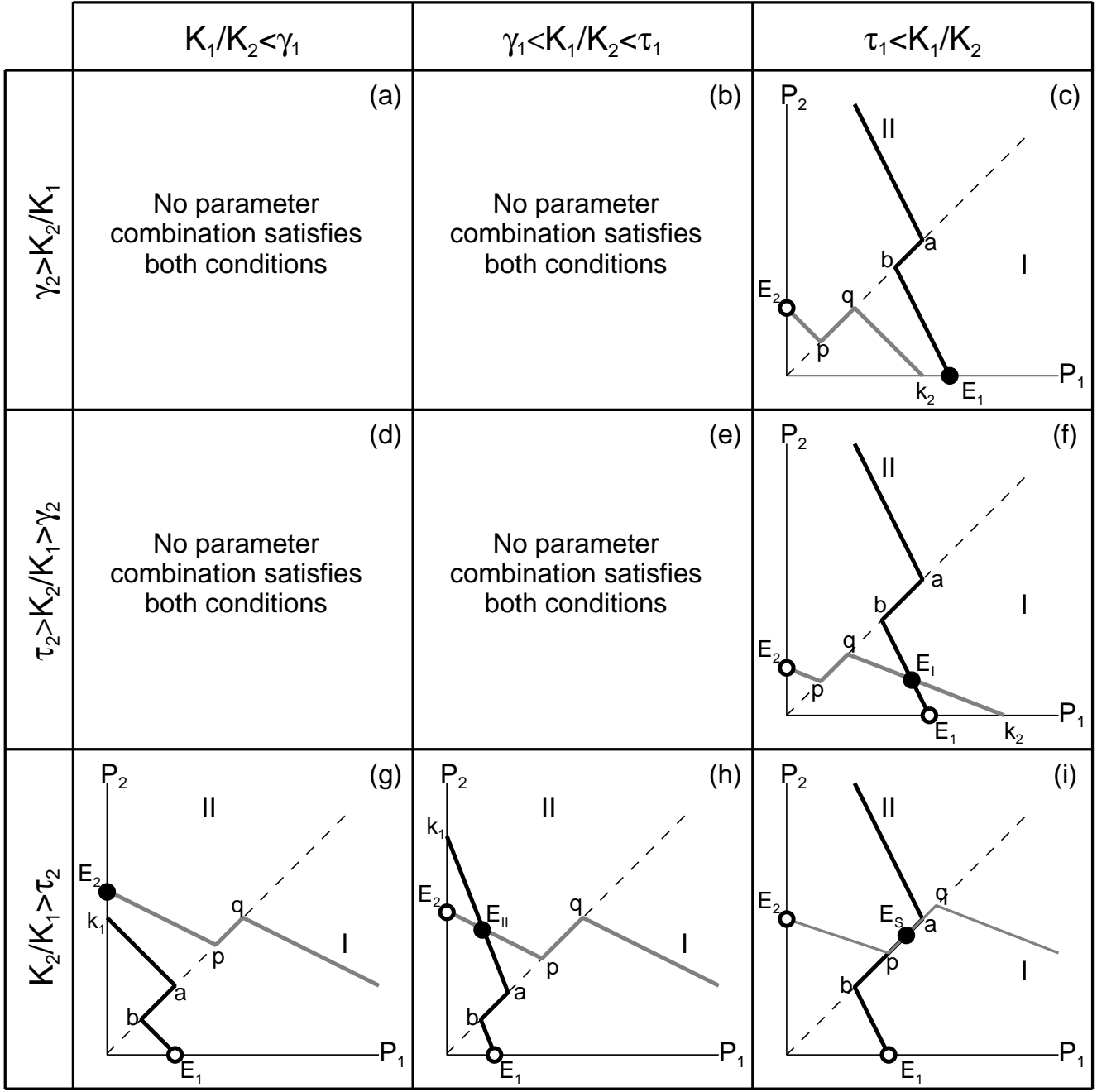} 
\par\end{centering}
\caption{\label{fig:partable_antweak}Plant generalized isoclines (plant 1: black, plant 2:
gray) under exploitation $(s=-1)$ and weak competition $(c_{1}c_{2}<1)$. Isoclines
intersect the dashed switching line (\ref{eq:A-switch}) at four points $\mathbf{a}$,
$\mathbf{b}$, $\mathbf{p}$, $\mathbf{q}$. Animals specialize on plant 1 (2) in
sector I (II) that is below (above) the switching line in the $P_{1}P_{2}$ plane.
Dots and circles denote stable and unstable equilibria (\ref{eq:E1},\ref{eq:E2},\ref{eq:E.I},\ref{eq:E.II},\ref{eq:E.S}),
respectively. Representative configurations are sketched according to carrying capacity
ratios in relation to invasion ($\gamma_{1}$, $\gamma_{2}$) and attraction ($\tau_{1}$,
$\tau_{2}$) thresholds.}
\end{figure}

All qualitatively different patterns of isoclines intersections when inter-specific
competition is weak and $A<r_{i}$ are shown in Figure \ref{fig:partable_antweak}.
Since $s=-1$, either $K_{1}/K_{2}>\tau_{1}$ or $K_{2}/K_{1}>\tau_{2}$, i.e., at
least one plant is always above its attraction threshold.\footnote{ The case where both $K_{1}/K_{2}<\tau_{1}$ and $K_{2}/K_{1}<\tau_{2}$ is not possible
because then $1<\tau_{1}\tau_{2}=(1-\frac{A}{r_{1}})(1-\frac{A}{r_{2}})<1$, a contradiction.} This is why Figure \ref{fig:partable_antweak}a, b, d, e are blank, because there
are no parameters that satisfy inequalities that define these four panels. With respect
to plant equilibria, there are three mutually exclusive possible outcomes of plant
competition.

First, the missing plant cannot invade the other plant monoculture equilibrium and
plant coexistence is not possible. These are situations where generalized isoclines
do not intersect nor overlap, and the dynamics globally converge toward the monoculture
equilibrium of the plant that can invade (to $\mathbf{E_{1}}$ in Figure \ref{fig:partable_antweak}c,
and to $\mathbf{E_{2}}$ in panel g).

Second, both plants can invade one another and the generalized isoclines intersect
in one of the two sectors. Thus, both plants coexist either at the globally stable
equilibrium $\mathbf{E_{I}}$ (panel f) at which exploiters specialize on plant 1,
or globally stable equilibrium $\mathbf{E_{II}}$ (panel h) at which exploiters specialize
on plant 2.

Third, generalized isoclines partially overlap along the switching line (Figure \ref{fig:partable_antweak}i),
so that there is globally stable equilibrium $\mathbf{E_{S}}$ at which animals behave
as generalists with intermediate preferences for plant 1 given by $\bar{u}_{1}$
in (\ref{eq:U.S}).

\subsection{Exploitation ($s=-1$) and strong inter-specific plant competition ($c_{1}c_{2}>1$)}

\begin{figure}
\begin{centering}
\includegraphics[width=1\textwidth]{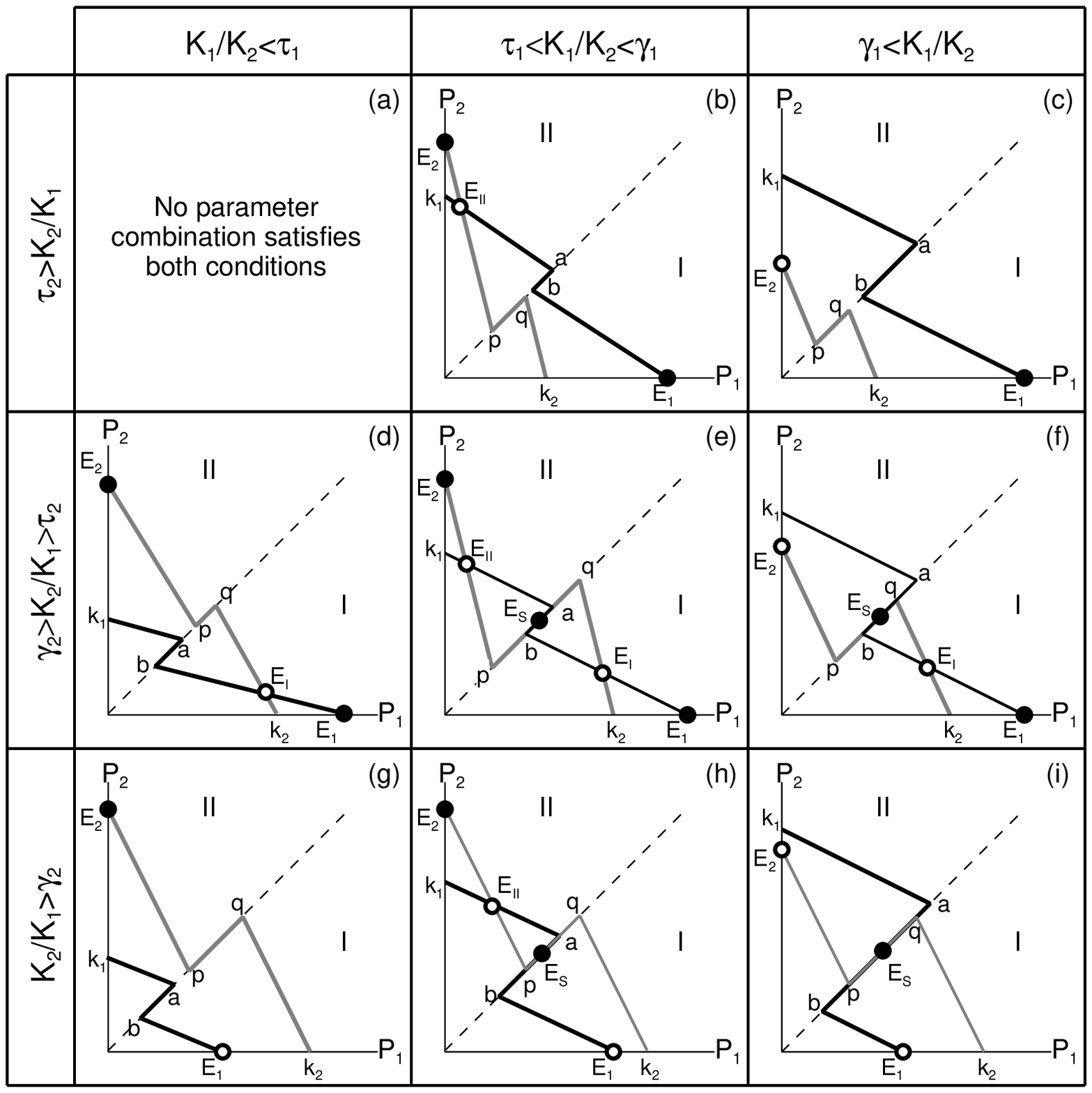} 
\par\end{centering}
\caption{\label{fig:partable_antstrong}Plant generalized isoclines under exploitation $(s=-1)$
and strong competition $(c_{1}c_{2}>1)$. Notation like in Figure \ref{fig:partable_antweak}.}
\end{figure}

Since $s=-1$, there are no parameters satisfying $K_{2}/K_{1}<\tau_{2}$ and $K_{1}/K_{2}<\tau_{1}$
exactly as in the previous case of weak competition and there are 8 qualitative cases
for isoclines intersections (Figure \ref{fig:partable_antstrong}).

Due to strong competition, stable plant coexistence is impossible in sector I or
sector II, but when both attraction thresholds are met (i.e., $K_{1}/K_{2}>\tau_{1}$
and $K_{2}/K_{1}>\tau_{2}$), the isoclines partially overlap along the switching
line and plants can coexist at equilibrium $\mathbf{E_{S}}$ where exploiters behave
as adaptive generalists with intermediate preference $\bar{u}_{1}$ for plant 1.
This state of coexistence can be locally or globally stable, depending on whether
invasion thresholds are met, as we will see next.

If neither of the two invasion thresholds are met (Figure \ref{fig:partable_antstrong}e),
equilibrium $\mathbf{E_{S}}$ is locally stable and depending on initial conditions
there are three possible outcomes for plant population dynamics: (i) monoculture
equilibrium $\mathbf{E_{1}}$ where exploiters specialize on plant 1 $(u_{1}=1)$
and plant 2 is excluded, (ii) monoculture equilibrium $\mathbf{E_{2}}$ where exploiters
specialize on plant 2 $(u_{1}=0)$ and plant 1 is excluded, or (iii) plant coexistence
equilibrium $\mathbf{E_{S}}$.

If only one plant invasion threshold is met, equilibrium $\mathbf{E_{S}}$ stays
locally stable and there is another monoculture equilibrium for the plant that meets
its invasion threshold (i.e., $\mathbf{E_{1}}$ in panel f, or $\mathbf{E_{2}}$
in panel h).

If both plants are above their invasion thresholds, $\mathbf{E_{S}}$ is globally
stable (Figure \ref{fig:partable_antstrong}i), despite of intra-specific competition
being stronger than inter-specific $(c_{1}c_{2}>1)$ that would not permit stable
coexistence in the standard LV competition model.

Like in standard LV models with strong competition, there are parameter values for
which generalized isoclines intersect in a single unstable equilibrium, leading to
the well known bi-stable outcome where plant 1 or plant 2 wins depending on initial
conditions (Figure \ref{fig:partable_antstrong}b,d).

\subsection{Mutualism ($s=1$) and weak inter-specific plant competition ($c_{1}c_{2}<1$)}

\label{section3.3} 
\begin{figure}
\begin{centering}
\includegraphics[width=1\textwidth]{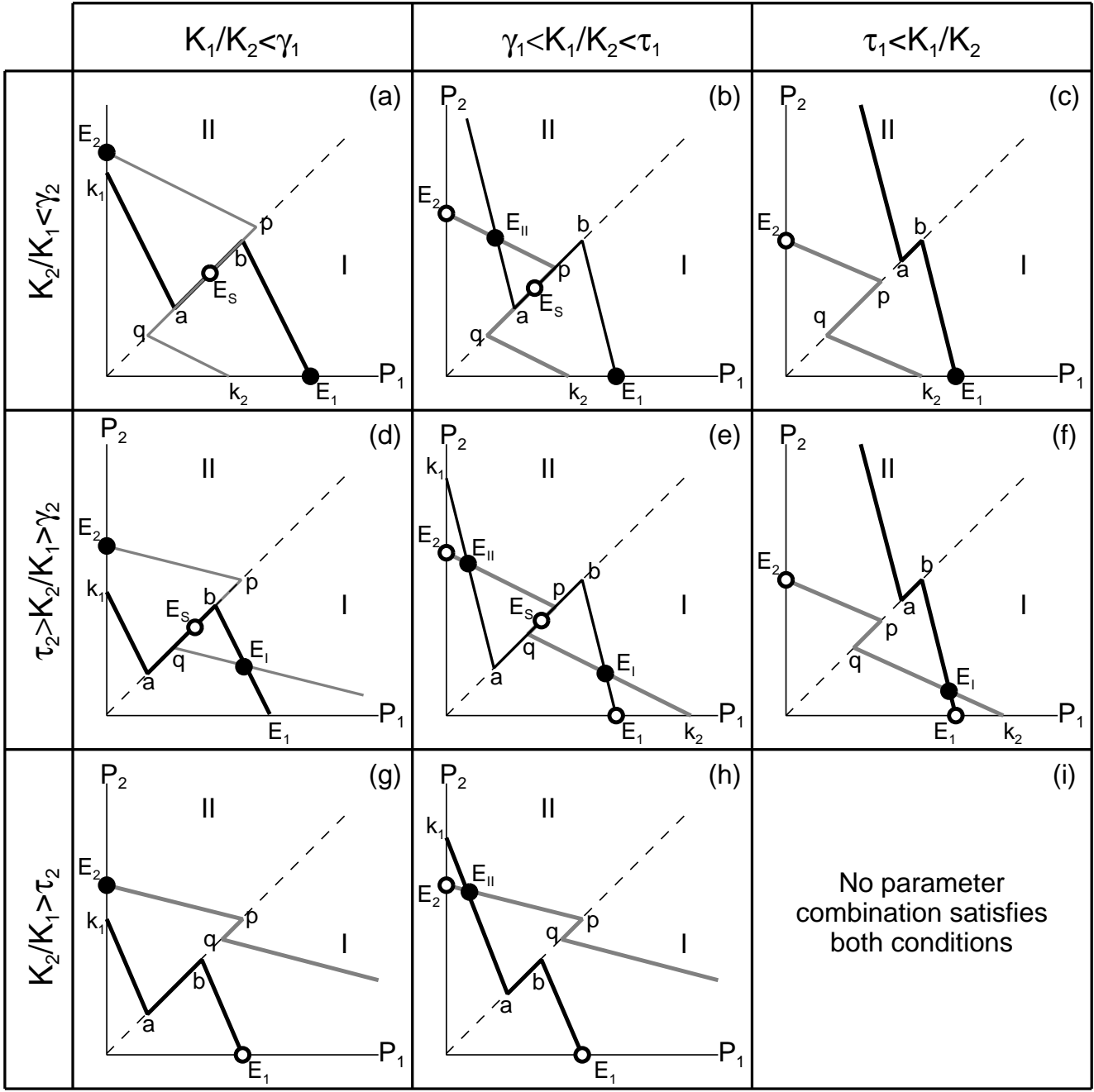} 
\par\end{centering}
\caption{\label{fig:partable_mutweak}Plant generalized isoclines under mutualism $(s=1)$
and weak competition $(c_{1}c_{2}<1)$. Notation like in Figure \ref{fig:partable_antweak}.}
\end{figure}

All possible qualitative intersections of isoclines under mutualism
and weak inter-specific plant competition are shown in Figure \ref{fig:partable_mutweak}.
As inter-specific competition is weak ($c_{1}c_{2}<1$), plant invasion thresholds
are smaller than attraction thresholds ($\gamma_{1}<\tau_{1}$ and $\gamma_{2}<\tau_{2}$)
and there are no parameter values such that $K_{1}/K_{2}>\tau_{1}$ and $K_{2}/K_{1}>\tau_{2}$,
i.e., panel i in Figure \ref{fig:partable_mutweak} is empty.\footnote{Indeed inequalities $K_{1}/K_{2}>\tau_{1}$ and $K_{2}/K_{1}>\tau_{2}$ imply that
$\tau_{1}\tau_{2}=(1+\frac{sA}{r_{1}})(1+\frac{sA}{r_{2}})<1$ which is false under
mutualism when $s=1$ .}

There are important differences in plant competition dynamics under mutualism when
compared to the exploitative case (cf.~Figure \ref{fig:partable_mutweak} vs.~Figure
\ref{fig:partable_antweak}). The main difference is that the interior equilibrium
$\mathbf{E_{S}}$, when it exists, is unstable for mutualism (Figure \ref{fig:partable_mutweak}a,
b, d, e). As this is the only plant coexistence equilibrium at which animals behave
as generalists, this predicts that mutualists will always behave as specialists when
plants are at a locally stable equilibrium, whether both plants coexist (Figure \ref{fig:partable_mutweak}b,
d, e, f, h) or not (Figure \ref{fig:partable_mutweak}a, c, g). The other important
difference between mutualists vs.~exploiters is that mutualism leads to alternative
locally stable plant equilibria (Figure \ref{fig:partable_mutweak}a, b, d, e). Where
the plant dynamics converge depends on initial plant population densities, and there
are three general cases that we describe next.

First, if neither plant invasion threshold is met (Figure \ref{fig:partable_mutweak}a)
initial conditions lead towards monoculture equilibrium $\mathbf{E_{1}}$ or $\mathbf{E_{2}}$,
where mutualists specialize on plant 1 or plant 2 respectively. This outcome is analogous
to the bi-stable case of the standard LV competition model when competition is strong
$(c_{1}c_{2}>1)$ and the interior equilibrium is a saddle point. But here, instead,
competition is weak $(c_{1}c_{2}<1)$, and bi-stability arises because equilibrium
$\mathbf{E_{S}}$ on the switching line behaves like a saddle point. We described
similar outcomes of mutual exclusion in previous obligatory mutualism models \citep{revilla_krivan-plosone16},
where plants competed exclusively for pollinator preferences (i.e., $c_{1}=c_{2}=0$).

Second, when plant 1 (2) meets its invasion threshold and the other plant 2 (1) does
not, initial conditions lead either to a monoculture of plant 1 (2) or to stable
coexistence of both plants with mutualists specializing on plant 2 (1) (e.g., $\mathbf{E_{1}}$
or $\mathbf{E_{II}}$ in Figure \ref{fig:partable_mutweak}b; $\mathbf{E_{2}}$ or
$\mathbf{E_{I}}$ in panel d).

Third, when both plants are above their invasion thresholds there are locally stable
equilibria in both sectors, and initial conditions determine whether coexistence
takes place at equilibrium $\mathbf{E_{I}}$ where mutualists specialize on plant
1, or at $\mathbf{E_{II}}$ where they specialize on plant 2 (Figure \ref{fig:partable_mutweak}e).

\subsection{Mutualism ($s=1$) and strong inter-specific plant competition ($c_{1}c_{2}>1$)}

\begin{figure}
\begin{centering}
\includegraphics[width=1\textwidth]{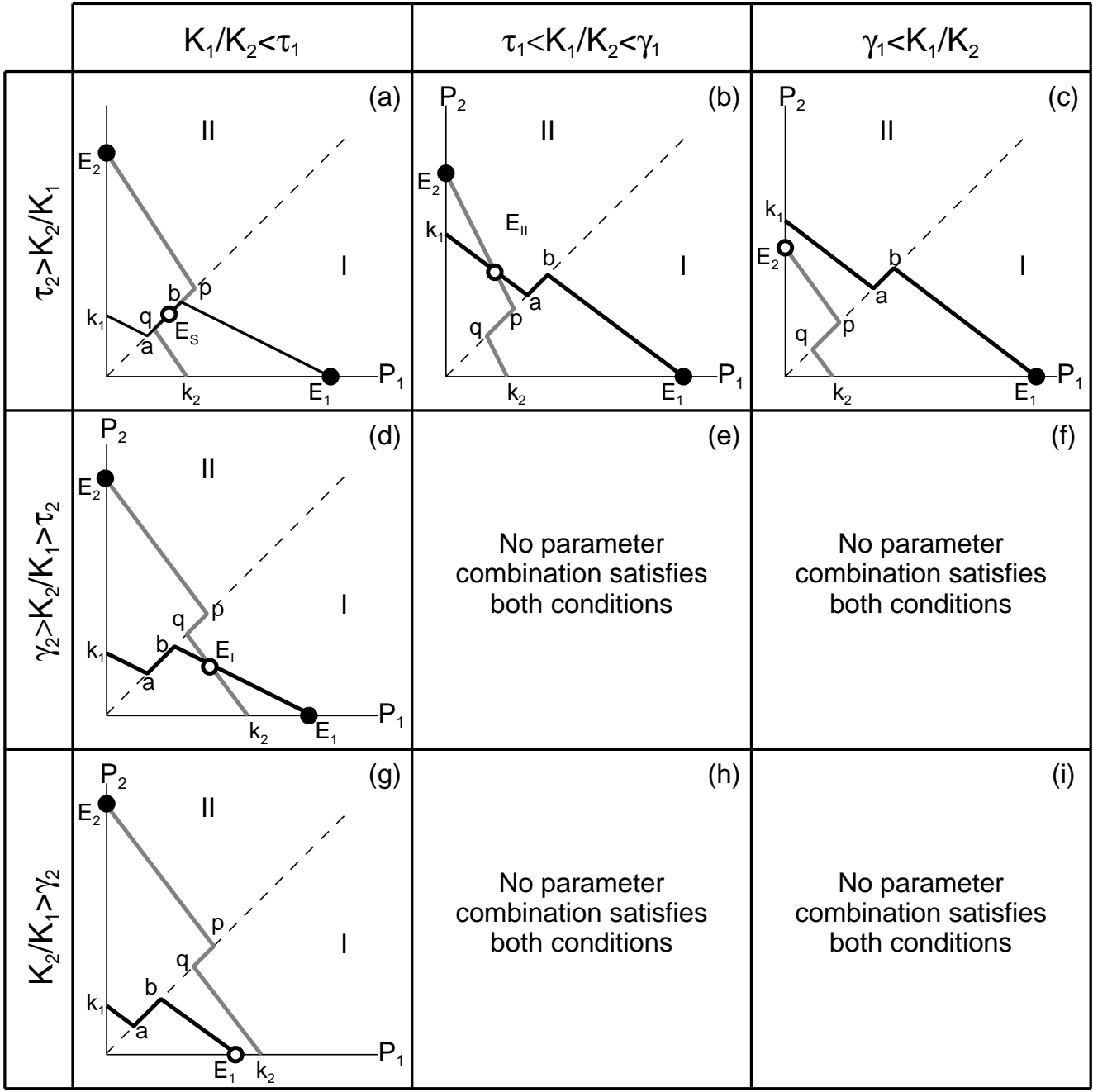} 
\par\end{centering}
\caption{\label{fig:partable_mutstrong}Plant generalized isoclines under mutualism $(s=1)$
and strong competition $(c_{1}c_{2}>1)$. Notation like in Figure \ref{fig:partable_antweak}.}
\end{figure}

When animals are mutualists ($s=1$) and inter-specific plant competition is strong
($c_{1}c_{2}>1$) attraction thresholds are smaller than invasion thresholds ($\gamma_{i}>\tau_{i},$
$i=1,2$) and there are no parameters satisfying $K_{2}/K_{1}>\gamma_{2}$ and $K_{1}/K_{2}>\gamma_{1}$
(i.e., panels e, f, h and i in Figure \ref{fig:partable_mutstrong} are empty). Moreover,
plant coexistence is impossible (Figure \ref{fig:partable_mutstrong}) which is in
a sharp contrast with the case of exploiters (Figure \ref{fig:partable_antstrong})
where plant coexistence is possible depending on initial conditions.

When isoclines intersect in sector I or II, and do not overlap along the switching
line, one plant competitively excludes the other plant, and plant population dynamics
are bi-stable (Figure \ref{fig:partable_mutstrong}b, d). These bi-stable scenarios
can be attributed entirely to strong inter-specific competition, like in standard
LV competition models. But again, as in the case of exploitation with strong competition,
bi-stability leads to mutualists specializing either on plant 1, or on plant 2. 

Bi-stability can also be caused by instability of equilibrium $\mathbf{E_{S}}$ when
the two plant isoclines partially overlap (Figure \ref{fig:partable_mutstrong}a),
similarly to the case where competition is weak as discussed in the previous Section
\ref{section3.3}.

\section{Discussion}

In this article we study effects of adaptive exploiters or mutualists on two competing
plant population dynamics, and on animal preference for plants. For plant population
dynamics described by the Lotka\textendash Volterra competition model we provide
a complete classification (Figures \ref{fig:partable_antweak}\textendash \ref{fig:partable_mutstrong},
Appendix \ref{sec:appclassification}) of coexistence states when plants interact
either with adaptive exploiters or mutualists that have fixed population densities.
This classification is based on comparing plant \emph{invasion } ($\gamma_{i}$ given
in (\ref{eq:coexSectorI})) and \emph{attraction} ($\tau_{i}$ given in (\ref{eq:coexSectorII}))
thresholds. These critical numbers capture the combined influences of (i) plant\textendash animal
interaction type (exploitation vs. mutualism), (ii) inter-specific plant competition
(weak vs.~strong), and (iii) indirect effects between plants mediated by changes
in animal preferences.

Model analysis leads to the following general predictions: 
\begin{enumerate}
\item Under exploitation and weak competition a globally stable plant coexistence equilibrium
exists when carrying capacities are not very unbalanced. At plant coexistence equilibrium
exploiters are specialist when at low densities while at high densities they are
generalists. Plant coexistence is possible even if neither of the two plants is viable
as a monoculture. 
\item Plant coexistence under exploitation and strong competition is possible but conditional,
i.e., depends on initial conditions. Up to three plant equilibria can co-exist. Plant
coexistence is possible only due to adaptive behavior of exploiters when exploiters
behave as adaptive generalists. 
\item Plant coexistence under mutualism and weak competition can be global or conditional
on initial plant population densities. Under mutualism animals always specialize
on the more profitable plant only. 
\item Plant coexistence under mutualism and strong competition is impossible. 
\end{enumerate}
An important special case when plants do not compete directly ($c_{1}=c_{2}=0$),
e.g., when plants grow in separate pots, or plants are separated by a fence or a
road \citep{geslin_etal-aer17}, was analyzed in \citet{krivan-tpb03} for exploiters.
In this case plant 1 (plant 2) isocline is vertical (horizontal) in sectors I and
II, invasion thresholds are zero so that they play no role at all, and attraction
thresholds simplify to $\tau_{1}=\frac{e_{2}}{e_{1}}\left(1+\frac{sA}{r_{2}}\right)$
and $\tau_{2}=\frac{e_{1}}{e_{2}}\left(1+\frac{sA}{r_{1}}\right)$. In the case of
exploitation $(s=-1)$ the only possible outcome is either global extinction (when
exploiter density is too high) or global coexistence where animals can be specialists
(when exploiter density is low) or generalists (when exploiter density is intermediate).
In the case of mutualism $(s=1)$ possible outcomes always predict coexistence, including
alternative stable states, as shown in Figure \ref{fig:partable_mutweak}e, f, h.

We stress here that our predictions concern a small community, and it would be incorrect
to extrapolate them to larger plant\textendash animal interaction networks without
proper consideration of model (\ref{eq:plant_odes}) limitations and assumptions
(see section Model assumption below). For example, our model predicts that both plants
can coexist with generalist exploiters but not with generalists mutualists, while
there is empirical evidence that insect pollinators are more generalist than insect
herbivores \citep{fontaine_etal-rspb09}. Disagreement arises, e.g., because our
Lotka\textendash Volterra model does not consider competition for plants among the
animals that are kept at fixed density. When mutualism is modeled under explicit
consumer\textendash resource dynamics where animal population densities change \citep{valdovinos_etal-oikos13,revilla_krivan-plosone16},
resource depletion (e.g., nectar consumption) can promote mutualist generalism, countering
the tendency towards exclusive specialization on the most profitable plant.

Another counter-intuitive prediction is that exploitation coupled with flexible foraging
enables indirect facilitation between plants. Once again, this is due to fixed animal
population densities, because increase in one plant population density does not lead
to increase in exploiter population density, a necessary condition for apparent competition
\citep{holt-tpb77} to occur. Even models that consider coupled prey\textendash predator
dynamics predict important positive effects between preys due to predator switching
\citep{abrams_matsuda-ecology96}. Thus by coexisting, plants share exploitation
costs, which leads to facilitation (i.e., higher equilibrium densities). Such indirect
positive effect can be extreme, i.e., plants that cannot tolerate exploitation alone
can survive when sharing exploitative stress with another plant (e.g., Figure \ref{fig:coex_with_expl}).
In the case of mutualism, flexible preferences gives rise to competition for preferences.
This increases competitive asymmetries already present between the plants (Figure
\ref{fig:coex_with_mut}). In addition, indirect facilitation between plants that
share mutualists \citep{waser_real-nature79} is prevented from happening by the
animals having fixed densities. In this respect, experiments show that competition
between plants for pollinator preferences can overcome such facilitation effects
\citep{ghazoul-joe06}.

\subsection*{Population dynamics and adaptive animal preferences}

To model effects of adaptive animal preferences on population dynamics of two competing
plant species, we combine the Lotka\textendash Volterra competitive model with a
behavioral model that describes changes in animal preferences for plants. This is
a common scenario in plant communities interacting with guilds of herbivores, parasites,
pollinators or seed dispersers \citep{melian_etal-oikos09,sauve_etal-theorecol15,bronstein2015}.
We assume that animal preferences for plants track instantaneously current plant
population densities which, in turn, influence plant population dynamics. To model
this feedback, we assume that animal preferences maximize animal fitness at current
plant population densities. As optimal animal preferences when both plants are equally
profitable are not uniquely given, the resulting plant population dynamics are described
by a Lotka\textendash Volterra differential inclusion \citep[e.g.,][]{colombo_krivan-jmamb93,krivan-tpb96,krivan-amnat97,krivan-amnat07,krivan_etal-tpb08}.
We analyze this model by generalizing the concept of isoclines which allows us to
provide a complete classification of all plant equilibria. To this end, we split
the plant phase space into two sectors (Figures \ref{fig:GIclines}a, c, e and \ref{fig:partable_antweak}\textendash \ref{fig:partable_mutstrong}).
The boundary that separates these sectors is called the switching line because animals
switch their preferences for plants when plant population numbers cross this line.
Along the switching line animal fitness is independent from animal preferences because
payoffs from both plants are the same. Inside the sectors, animals specialize on
one plant only\footnote{Similar concepts, called isodars and isolegs, are used in the habitat selection theory
\citep{pimm_rosenzweig-oikos81,rosenzweig-ecology81,krivan_sirot-amnat02,morris-oecologia03,krivan_vrkoc-jmb07}
where distribution of a single population is studied as a function of the number
of individuals of that population. In this article distribution of animals depends
not only on animal population density, but also on plant densities.}. Thus, plant isoclines inside each sector coincide with the classical isoclines
for the Lotka\textendash Volterra competition model. In this article we define generalized
plant isoclines that are formed by sector-wise pieces of isoclines that are connected
with segments of the switching line (Figures \ref{fig:partable_antweak}\textendash \ref{fig:partable_mutstrong}).
Thus, globally, generalized plant isoclines are piece-wise linear, which leads to
multiple isocline intersections and multiple steady states. In particular, the segments
of the two plant isoclines that are subsets of the switching line can partially overlap
(Figure \ref{fig:partable_antweak}i; Figure \ref{fig:partable_antstrong}e, f, h,
i; Figure \ref{fig:partable_mutweak}a, b, d, e; Figure \ref{fig:partable_mutstrong}a).
If so, we show that plant population dynamics have a unique equilibrium in this overlapping
segment (e.g., Figure \ref{fig:GIclines}b, c). This equilibrium is either locally
stable when animals are exploiters or unstable when animals are mutualists. There
are important differences between plant equilibria in the switching line and those
that are inside sectors because animals are specialists inside sectors, but they
are generalists at the equilibrium that is in the switching line.

The configuration of generalized isoclines depends on plant \emph{invasion thresholds}
(\ref{eq:coexSectorI}) and \emph{attraction thresholds} (\ref{eq:coexSectorII}).
Invasion thresholds $\gamma_{i}$ determine whether the missing plant species can
invade the other plant monoculture at the equilibrium. For the standard Lotka\textendash Volterra
competition model with fixed animal preferences, coexistence as well as global dynamics
can be predicted entirely in terms of invasion thresholds. However, when interactions
between plants and animals are adaptive, we have to consider animal preferences which
leads to non-linear generalized isoclines, and the concept of attraction thresholds.
Attraction threshold $\tau_{i}$ determines whether the plant coexistence equilibrium
at the switching line, where animals behave as plant generalists, locally attracts
or repels orbits. This is analogous to the invasion threshold which determines whether
the boundary equilibria attract or repel orbits. Attraction thresholds depend on
animal density, inter-specific competition, and on payoffs ($e_{i}$) animals obtain
from plants. These payoffs define animal fitness which is a function of plant densities.
Despite the fact that we assume fixed animal densities, animal preferences (i.e.,
animal behavioral traits) change with changes in plant numbers. In other words, we
observe indirect interactions between plants mediated by changes in animal preferences
\citep[i.e., trait-mediated indirect interactions between plants sensu ][]{bolker_etal-ecology03}.
Thus, attraction thresholds capture the combined effects of inter-specific plant
competition and behaviorally-mediated indirect effects, and their positions relative
to invasion thresholds determine global interaction dynamics as summarized at the
start of the discussion section.

\subsection*{Model assumptions}

The plant\textendash animal model assumes constant animal density. This allows us
to focus on behavior-mediated indirect interactions between plants not affected by
simultaneous density-mediated interactions caused by changes in animal density (i.e.,
apparent competition and facilitation). This is reasonable assumption if animal populations
are regulated mainly by external factors not explicitly considered. A good example
is the case of common bees with large managed populations \citep{geslin_etal-aer17},
spilling over natural communities. Constant animal density is also enforced in short
term experiments that study the effect of foraging behavior on plant success \citep{fontaine_etal-plosbiol05}.
Another plausible scenario is that the animal population dynamics is very slow when
compared with plants due to differences in generation time (e.g., ungulate recruitment
being slower than grass regrowth). An important prediction of the model is that exploitation
favors animal generalism, while mutualism favors specialization. When animal population
dynamics are considered, animal benefits must decrease due to intra-specific competition
for plant resources, favoring generalism over specialization, even under mutualism
\citep{revilla_krivan-plosone16}.

Another important assumption is that animal adaptation is much faster than plant
population dynamics. This requires that changes of foraging behavior occur within
individual lifetimes, e.g., highly mobile consumers dispersing between plant species,
like in the ideal free distribution \citep{krivan-tpb03}. The assumption of fast
adaptation can be relaxed by modeling preference dynamics explicitly using, e.g.,
replicator equation \citep{kondoh-science03}. In \citet{revilla_krivan-plosone16}
we showed that qualitative predictions related to mutualist generalism vs. specialism
are preserved even when adaptation runs on a similar time scale as population dynamics.
However, when adaptation was much slower than population dynamics, predictions frequently
diverged due to extreme dependence on animal initial preferences. For example, if
animals initially strongly prefer one plant over the other despite the fact that
such behavior is not optimal, the initially preferred plant can die out before animal
preferences could change. In addition, when adaptation occurs over multiple generations,
specialization or generalism also depends on the evolution of fitness related traits
such as conversion efficiencies ($e_{i}$), which scale interactions with payoffs.
Parameters like these depend on complex morphological and physiological constraints,
and they generally relate to one another via non-linear trade-offs \citep{egas_etal-amnat04}.
Accounting for long term change of these parameters requires different approaches
(e.g., adaptive dynamics, \citealt{kisdi-amnat02,egas_etal-amnat04,rueffler_etal-amnat06}).

Finally, we only consider facultative mutualism because many plants have multiple
pollinators or seed dispersers \citep{melian_etal-oikos09}. Obligate mutualism can
be modeled with Lotka\textendash Volterra equations \citep{vandermeer_boucher-jtb78},
but adaptive preference rules out coexistence trivially because mutualists interact
with the more profitable plant only and the less profitable plant will die. Obligate
mutualisms are better studied using mechanistic models \citep{revilla_krivan-plosone16,revilla_krivan-jtb18},
that predict coexistence depending on initial conditions because of mutualistic Allee
effects \citep{bronstein2015}.

\subsection*{Conditional coexistence and alternative plant stable states}

The interplay between plant competition and animal adaptation gives rise to complex
plant population\textendash animal preference dynamics. As plant isoclines are non-linear
(e.g., Figure \ref{eq:isoclines}) multiple equilibria can co-exist. This has important
implications for the diversity of communities under perturbations \citep{yan_zhang-rspb14,zhang_etal-ecomod15}.
On the one hand, perturbations in plant abundances can lead to loss of coexistence
under exploitation and strong competition, i.e. coexistence conditioned by initial
conditions (e.g., Figure \ref{fig:partable_antstrong}e, f, h). On the other hand,
they can trigger transitions between alternative stable states of coexistence when
mutualism and weak inter-plant competition combine (e.g., Figure \ref{fig:partable_mutweak}e).

Conditional coexistence and coexistence at alternative stable states are common predictions
of models that combine positive and negative density-dependent interactions \citep[e.g.,][]{hernandez-rspb98,holland_deangelis-ecology10,revilla_encinas-plosone15}.
In the present model, however, plants always interact negatively due to inter-specific
competition, and additional positive or negative effects arise due to adaptive preference
of common exploiters or mutualists. Since animal densities are fixed, these indirect
effects are behavior-mediated, but triggered by changes in plant densities. It is
very important to remark that such abundance\textendash preference feedbacks between
trophic levels leads to very different predictions when compared to abundance\textendash abundance
feedbacks between trophic levels. In this latter case where animals respond numerically
to plant densities, exploitation leads to \emph{apparent competition} \citep{holt-tpb77}
and mutualism to \emph{apparent mutualism} (or apparent facilitation) between plants,
which respectively opposes and favors coexistence \citep{sauve_etal-theorecol15}.
When animal preferences respond to plant densities, exploitation leads to a \emph{competitive
release} that promotes stable plant coexistence \citep{krivan-amnat97,krivan-eer03}
while mutualism leads to \emph{competition for mutualists preferences} between plants
that destabilizes plant coexistence and leads to plant exclusion \citep{revilla_krivan-jtb18}.

In this article we showed that conditional plant coexistence is expected in scenarios
where generalist exploiters regulate strongly competing plants (i.e., $c_{1}c_{2}>1$,
Figure \ref{fig:partable_antstrong}e, f, h, i). On the other hand, outcomes like
coexistence at alternative stable states are expected between weakly competing plants
(i.e., $c_{1}c_{2}<1$) that are regulated by specialized mutualists. How relevant
these predictions are in the real world depends on how widespread are situations
where intra-specific competition is stronger than inter-specific, and vice-versa.
On the one hand, meta-analyses suggest that intra- and inter-specific effects are
too similar to be discerned \citep{gurevitch_etal-amnat92}, or that intra-specific
effects are actually much stronger than inter-specific (i.e., $c_{1}c_{2}\leq1$;
\citealp{adler_etal-ecolett18}). However, recent pair-wise competition experiments
\citep{sheppard-bionv19} suggest that inter-specific competition can be strong (i.e.,
$c_{1}c_{2}>1$). Such uncertainty is rooted in the fact that these surveys assume
models like (\ref{eq:plant_odes}) that treat competition phenomenologically, and
there can be multiple underlying factors that can lead to strong net competition.
For example, competition can be strengthened by allelopathy \citep{inderjit_delmoral-tbr97},
which is decidedly stronger against non-specifics compared to con-specifics.

It will be interesting to explore to what extent our conclusions can be extrapolated
to larger communities, consisting of several animal and plant species. For such diverse
scenarios coexistence must result from intricate balances between multiple positive
and negative effects \citep{melian_etal-oikos09,georgelin_loeuille-jtb14,mougi_kondoh-ecores14,revilla_krivan-plosone16},
where density- and behaviorally-mediated effects mix up. The analytical study of
combined exploitative and mutualist effects is more difficult. For an illustration,
let us consider a second exploiter or mutualist. This modification of model (\ref{eq:plant_odes})
will result in two switching lines (one for each animal species), three sectors and
piece-wise continuous generalized isoclines that will consist of five segments. Mathematical
analysis given in this article can be extended to describe this case as well, but
as the number of species increases, complete mathematical classification will be
intractable due to combinatorial complexity of possible outcomes. In these cases
simulation approaches can be useful for studying the likelihood of multiple equilibria,
as a function of competition intensity and the proportion of exploitative vs. mutualistic
interactions (e.g., \citealt{melian_etal-oikos09}).

\section*{Acknowledgements}

This project has received funding from the European Union's Horizon 2020 research
and innovation programme under the Marie Sk\l odowska-Curie grant agreement No 690817.
We also thank the two referees and the Handling Editor for their encouraging comments
on the original submission.

\bibliographystyle{chicago}
\bibliography{pla_exp_mut}

\begin{thebibliography}{}

\bibitem[\protect\citeauthoryear{Abrams and Matsuda}{Abrams and
  Matsuda}{1996}]{abrams_matsuda-ecology96}
Abrams, P.~A. and H.~Matsuda (1996).
\newblock {P}ositive indirect effects between prey species that share
  predators.
\newblock {\em {E}cology\/}~{\em 77}, 610--616.

\bibitem[\protect\citeauthoryear{Adler, Smull, Beard, Choi, Furniss,
  Kulmatiski, Meiners, Tredennick, and Veblen}{Adler
  et~al.}{2018}]{adler_etal-ecolett18}
Adler, P.~B., D.~Smull, K.~H. Beard, R.~T. Choi, T.~Furniss, A.~Kulmatiski,
  J.~M. Meiners, A.~T. Tredennick, and K.~E. Veblen (2018).
\newblock {C}ompetition and coexistence in plant communities: intraspecific
  competition is stronger than interspecific competition.
\newblock {\em {E}cology {L}etters\/}~{\em 21}, 1319--1329.

\bibitem[\protect\citeauthoryear{Aubin and Cellina}{Aubin and
  Cellina}{1984}]{aubincellina1984}
Aubin, J.~P. and A.~Cellina (1984).
\newblock {\em {D}ifferential {I}nclusions: {S}et-valued {M}aps and {V}iability
  {T}heory}.
\newblock Springer-Verlag.

\bibitem[\protect\citeauthoryear{Bastolla, Fortuna, Pascual-García, Ferrera,
  Luque, and Bascompte}{Bastolla et~al.}{2009}]{bastolla_etal-nature09}
Bastolla, U., M.~A. Fortuna, A.~Pascual-García, A.~Ferrera, B.~Luque, and
  J.~Bascompte (2009).
\newblock {T}he architecture of mutualistic networks minimizes competition and
  increases biodiversity.
\newblock {\em {N}ature\/}~{\em 458}, 1018--1020.

\bibitem[\protect\citeauthoryear{Berec, Eisner, and Krivan}{Berec
  et~al.}{2010}]{berec_etal-jtb10}
Berec, L., J.~Eisner, and V.~Krivan (2010).
\newblock {A}daptive foraging does not always lead to more complex food webs.
\newblock {\em {J}ournal of {T}heoretical {B}iology\/}~{\em 266}, 211--218.

\bibitem[\protect\citeauthoryear{Bolker, Holyoak, K\v{r}ivan, Rowe, and
  Schmitz}{Bolker et~al.}{2003}]{bolker_etal-ecology03}
Bolker, B., M.~Holyoak, V.~K\v{r}ivan, L.~Rowe, and O.~Schmitz (2003).
\newblock {C}onnecting theoretical and empirical studies of trait-mediated
  interactions.
\newblock {\em {E}cology\/}~{\em 84}, 1101--1114.

\bibitem[\protect\citeauthoryear{Bronstein}{Bronstein}{2015}]{bronstein2015}
Bronstein, J.~L. (2015).
\newblock {\em {M}utualism}.
\newblock Oxford University Press.

\bibitem[\protect\citeauthoryear{Case}{Case}{2000}]{case2000}
Case, T. (2000).
\newblock {\em {A}n {I}llustrated {G}uide to {T}heoretical {E}cology}.
\newblock Oxford University Press.

\bibitem[\protect\citeauthoryear{Colombo and K\v{r}ivan}{Colombo and
  K\v{r}ivan}{1993}]{colombo_krivan-jmamb93}
Colombo, R. and V.~K\v{r}ivan (1993).
\newblock {S}elective strategies in food webs.
\newblock {\em {IMA} {J}ournal of {M}athematics {A}pplied in {M}edicine and
  {B}iology\/}~{\em 10}, 281--291.

\bibitem[\protect\citeauthoryear{Egas, Dieckmann, and Sabelis}{Egas
  et~al.}{2004}]{egas_etal-amnat04}
Egas, M., U.~Dieckmann, and M.~W. Sabelis (2004).
\newblock {E}volution restricts the coexistence of specialists and generalists:
  the role of trade-off structure.
\newblock {\em {A}merican {N}aturalist\/}~{\em 163}, 518--531.

\bibitem[\protect\citeauthoryear{Feinsinger}{Feinsinger}{1987}]{feinsinger-tree87}
Feinsinger, P. (1987).
\newblock {E}ffects of plant species on each other's pollination: is community
  structure influenced?
\newblock {\em {T}rends in {E}cology and {E}volution\/}~{\em 2}, 123--126.

\bibitem[\protect\citeauthoryear{Filippov}{Filippov}{1988}]{filippov1988}
Filippov, A.~F. (1988).
\newblock {\em {D}ifferential equations with discontinuous righthand sides}.
\newblock Academic Publishers.

\bibitem[\protect\citeauthoryear{Fontaine, Dajoz, Meriguet, and
  Loreau}{Fontaine et~al.}{2005}]{fontaine_etal-plosbiol05}
Fontaine, C., I.~Dajoz, J.~Meriguet, and M.~Loreau (2005).
\newblock {F}unctional diversity of plant--pollinator interaction webs enhances
  the persistence of plant communities.
\newblock {\em {PL}o{S} {B}iology\/}~{\em 4\/}(1), e1.

\bibitem[\protect\citeauthoryear{Fontaine, Th{\'e}bault, and Dajoz}{Fontaine
  et~al.}{2009}]{fontaine_etal-rspb09}
Fontaine, C., E.~Th{\'e}bault, and I.~Dajoz (2009).
\newblock {A}re insect pollinators more generalist than insect herbivores?
\newblock {\em {P}roceedings of the {R}oyal {S}ociety {B}: {B}iological
  {S}ciences\/}~{\em 276}, 3027--3033.

\bibitem[\protect\citeauthoryear{Gause}{Gause}{1934}]{gause1934}
Gause, G.~F. (1934).
\newblock {\em {T}he {S}truggle for {E}xistence}.
\newblock Baltimore, MD: Williams \& Wilkins.

\bibitem[\protect\citeauthoryear{Georgelin and Loeuille}{Georgelin and
  Loeuille}{2014}]{georgelin_loeuille-jtb14}
Georgelin, E. and N.~Loeuille (2014).
\newblock {D}ynamics of coupled mutualistic and antagonistic interactions, and
  their implications for ecosystem management.
\newblock {\em {J}ournal of {T}heoretical {B}iology\/}~{\em 346}, 67--74.

\bibitem[\protect\citeauthoryear{Geslin, Gauzens, Baude, Dajoz, Fontaine,
  Henry, Ropars, Rollin, Th{\'e}bault, and Vereecken}{Geslin
  et~al.}{2017}]{geslin_etal-aer17}
Geslin, B., B.~Gauzens, M.~Baude, I.~Dajoz, C.~Fontaine, M.~Henry, L.~Ropars,
  O.~Rollin, E.~Th{\'e}bault, and N.~Vereecken (2017).
\newblock {M}assively {I}ntroduced {M}anaged {S}pecies and {T}heir
  {C}onsequences for {P}lant--{P}ollinator {I}nteractions.
\newblock {\em {A}dvances in {E}cological {R}esearch\/}~{\em 57}, 147--199.

\bibitem[\protect\citeauthoryear{Ghazoul}{Ghazoul}{2006}]{ghazoul-joe06}
Ghazoul, J. (2006).
\newblock {F}loral diversity and the facilitation of pollination.
\newblock {\em {J}ournal of {E}cology\/}~{\em 94}, 295--304.

\bibitem[\protect\citeauthoryear{Grover}{Grover}{1997}]{grover1997}
Grover, J.~P. (1997).
\newblock {\em {R}esource {C}ompetition}.
\newblock Chapman \& Hall.

\bibitem[\protect\citeauthoryear{Gurevitch, Morrow, Wallace, and
  Walsh}{Gurevitch et~al.}{1992}]{gurevitch_etal-amnat92}
Gurevitch, J., L.~L. Morrow, A.~Wallace, and J.~S. Walsh (1992).
\newblock {A} meta-analysis of competition in field experiments.
\newblock {\em {A}merican {N}aturalist\/}~{\em 140}, 539--572.

\bibitem[\protect\citeauthoryear{Hardin}{Hardin}{1960}]{hardin-science60}
Hardin, G. (1960).
\newblock {T}he competitive exclusion principle.
\newblock {\em {S}cience\/}~{\em 131}, 1292--1297.

\bibitem[\protect\citeauthoryear{Hernandez}{Hernandez}{1998}]{hernandez-rspb98}
Hernandez, M.~J. (1998).
\newblock {D}ynamics of transitions between population interactions: a
  nonlinear interaction $\alpha$-function defined.
\newblock {\em {P}roceedings of the {R}oyal {S}ociety {B}: {B}iological
  {S}ciences\/}~{\em 265}, 1433--1440.

\bibitem[\protect\citeauthoryear{Holland and DeAngelis}{Holland and
  DeAngelis}{2010}]{holland_deangelis-ecology10}
Holland, J.~N. and D.~L. DeAngelis (2010).
\newblock {A} consumer-resource approach to the density-dependent population
  dynamics of mutualism.
\newblock {\em {E}cology\/}~{\em 91}, 1286--1295.

\bibitem[\protect\citeauthoryear{Holt}{Holt}{1977}]{holt-tpb77}
Holt, R.~D. (1977).
\newblock {P}redation, apparent competition, and the structure of prey
  communities.
\newblock {\em {T}heoretical {P}opulation {B}iology\/}~{\em 12}, 197--229.

\bibitem[\protect\citeauthoryear{Holt, Grover, and Tilman}{Holt
  et~al.}{1994}]{holt_etal-amnat94}
Holt, R.~D., J.~Grover, and D.~Tilman (1994).
\newblock {S}imple rules for interspecific dominance in systems with
  exploitative and apparent competition.
\newblock {\em {A}merican {N}aturalist\/}~{\em 144}, 741--771.

\bibitem[\protect\citeauthoryear{Inderjit and Del~Moral}{Inderjit and
  Del~Moral}{1997}]{inderjit_delmoral-tbr97}
Inderjit and R.~Del~Moral (1997).
\newblock {I}s separating resource competition from allelopathy realistic?
\newblock {\em {T}he {B}otanical {R}eview\/}~{\em 63}, 221--230.

\bibitem[\protect\citeauthoryear{Kisdi}{Kisdi}{2002}]{kisdi-amnat02}
Kisdi, {\'E}. (2002).
\newblock {D}ispersal: risk spreading versus local adaptation.
\newblock {\em {A}merican {N}aturalist\/}~{\em 159}, 579--596.

\bibitem[\protect\citeauthoryear{Kondoh}{Kondoh}{2003}]{kondoh-science03}
Kondoh, M. (2003).
\newblock {F}oraging adaptation and the relationship between food-web
  complexity and stability.
\newblock {\em {S}cience\/}~{\em 299}, 1388--1391.

\bibitem[\protect\citeauthoryear{K\v{r}ivan}{K\v{r}ivan}{1996}]{krivan-tpb96}
K\v{r}ivan, V. (1996).
\newblock {O}ptimal foraging and predator prey dynamics.
\newblock {\em {T}heoretical {P}opulation {B}iology\/}~{\em 49}, 265--290.

\bibitem[\protect\citeauthoryear{K\v{r}ivan}{K\v{r}ivan}{1997}]{krivan-amnat97}
K\v{r}ivan, V. (1997).
\newblock {D}ynamic ideal free distribution: effects of optimal patch choice on
  predator-prey dynamics.
\newblock {\em {A}merican {N}aturalist\/}~{\em 149}, 164--178.

\bibitem[\protect\citeauthoryear{K\v{r}ivan}{K\v{r}ivan}{2003a}]{krivan-eer03}
K\v{r}ivan, V. (2003a).
\newblock {C}ompetitive co-existence caused by adaptive predators.
\newblock {\em {E}volutionary {E}cology {R}esearch\/}~{\em 5}, 1163--1182.

\bibitem[\protect\citeauthoryear{K\v{r}ivan}{K\v{r}ivan}{2003b}]{krivan-tpb03}
K\v{r}ivan, V. (2003b).
\newblock {I}deal free distributions when resources undergo population
  dynamics.
\newblock {\em {T}heoretical {P}opulation {B}iology\/}~{\em 64}, 25--38.

\bibitem[\protect\citeauthoryear{K\v{r}ivan}{K\v{r}ivan}{2007}]{krivan-amnat07}
K\v{r}ivan, V. (2007).
\newblock {T}he {L}otka--{V}olterra predator-prey model with
  foraging--predation risk trade-offs.
\newblock {\em {A}merican {N}aturalist\/}~{\em 170}, 771--782.

\bibitem[\protect\citeauthoryear{K\v{r}ivan}{K\v{r}ivan}{2010}]{krivan-jtb10}
K\v{r}ivan, V. (2010).
\newblock {E}volutionary stability of optimal foraging: partial preferences in
  the diet and patch models.
\newblock {\em {J}ournal of {T}heoretical {B}iology\/}~{\em 267}, 486--494.

\bibitem[\protect\citeauthoryear{K\v{r}ivan, Cressman, and
  Schneider}{K\v{r}ivan et~al.}{2008}]{krivan_etal-tpb08}
K\v{r}ivan, V., R.~Cressman, and C.~Schneider (2008).
\newblock {T}he ideal free distribution: a review and synthesis of the
  game-theoretic perspective.
\newblock {\em {T}heoretical {P}opulation {B}iology\/}~{\em 73}, 403--425.

\bibitem[\protect\citeauthoryear{K\v{r}ivan and Schmitz}{K\v{r}ivan and
  Schmitz}{2004}]{krivan_schmitz-oikos04}
K\v{r}ivan, V. and O.~J. Schmitz (2004).
\newblock {T}rait and density mediated indirect interactions in simple food
  webs.
\newblock {\em {O}ikos\/}~{\em 107}, 239--250.

\bibitem[\protect\citeauthoryear{K\v{r}ivan and Sirot}{K\v{r}ivan and
  Sirot}{2002}]{krivan_sirot-amnat02}
K\v{r}ivan, V. and E.~Sirot (2002).
\newblock {H}abitat selection by two competing species in a two-habitat
  environment.
\newblock {\em {A}merican {N}aturalist\/}~{\em 160}, 214--234.

\bibitem[\protect\citeauthoryear{K\v{r}ivan and Vrko\v{c}}{K\v{r}ivan and
  Vrko\v{c}}{2007}]{krivan_vrkoc-jmb07}
K\v{r}ivan, V. and I.~Vrko\v{c} (2007).
\newblock {A} {L}yapunov function for piecewise-independent differential
  equations: stability of the ideal free distribution in two patch
  environments.
\newblock {\em {J}ournal of {M}athematical {B}iology\/}~{\em 54}, 465--488.

\bibitem[\protect\citeauthoryear{MacArthur and Levins}{MacArthur and
  Levins}{1967}]{macarthur_levins-amnat67}
MacArthur, R.~H. and R.~Levins (1967).
\newblock {T}he limiting similarity, convergence, and divergence of coexisting
  species.
\newblock {\em {A}merican {N}aturalist\/}~{\em 101}, 377--385.

\bibitem[\protect\citeauthoryear{Melián, Bascompte, Jordano, and
  K\v{r}ivan}{Melián et~al.}{2009}]{melian_etal-oikos09}
Melián, C.~J., J.~Bascompte, P.~Jordano, and V.~K\v{r}ivan (2009).
\newblock {D}iversity in a complex ecological network with two interaction
  types.
\newblock {\em {O}ikos\/}~{\em 118}, 122--130.

\bibitem[\protect\citeauthoryear{Morris}{Morris}{2003}]{morris-oecologia03}
Morris, D.~W. (2003).
\newblock {T}oward an ecological synthesis: a case for habitat selection.
\newblock {\em {O}ecologia\/}~{\em 136}, 1--13.

\bibitem[\protect\citeauthoryear{Mougi and Kondoh}{Mougi and
  Kondoh}{2014}]{mougi_kondoh-ecores14}
Mougi, A. and M.~Kondoh (2014).
\newblock {A}daptation in a hybrid world with multiple interaction types: a new
  mechanism for species coexistence.
\newblock {\em {E}cological {R}esearch\/}~{\em 29}, 113--119.

\bibitem[\protect\citeauthoryear{Pimm and Rosenzweig}{Pimm and
  Rosenzweig}{1981}]{pimm_rosenzweig-oikos81}
Pimm, S.~L. and M.~L. Rosenzweig (1981).
\newblock {C}ompetitors and habitat use.
\newblock {\em {O}ikos\/}~{\em 37}, 1--6.

\bibitem[\protect\citeauthoryear{Revilla and Encinas-Viso}{Revilla and
  Encinas-Viso}{2015}]{revilla_encinas-plosone15}
Revilla, T.~A. and F.~Encinas-Viso (2015).
\newblock {D}ynamical transitions in a pollination--herbivory interaction: a
  conflict between mutualism and antagonism.
\newblock {\em {PL}o{S} {ONE}\/}~{\em 10}, e0117964.

\bibitem[\protect\citeauthoryear{Revilla and K\v{r}ivan}{Revilla and
  K\v{r}ivan}{2016}]{revilla_krivan-plosone16}
Revilla, T.~A. and V.~K\v{r}ivan (2016).
\newblock {P}ollinator foraging adaptation and the coexistence of competing
  plants.
\newblock {\em {PL}o{S} {ONE}\/}~{\em 11}, e0160076.

\bibitem[\protect\citeauthoryear{Revilla and K\v{r}ivan}{Revilla and
  K\v{r}ivan}{2018}]{revilla_krivan-jtb18}
Revilla, T.~A. and V.~K\v{r}ivan (2018).
\newblock {C}ompetition, trait-mediated facilitation, and the structure of
  plant-pollinator communities.
\newblock {\em {J}ournal of {T}heoretical {B}iology\/}~{\em 440}, 42--57.

\bibitem[\protect\citeauthoryear{Rohr, Saavedra, and Bascompte}{Rohr
  et~al.}{2014}]{rohr_etal-science14}
Rohr, R.~P., S.~Saavedra, and J.~Bascompte (2014).
\newblock {O}n the structural stability of mutualistic systems.
\newblock {\em {S}cience\/}~{\em 345}, 1253497.

\bibitem[\protect\citeauthoryear{Rosenzweig}{Rosenzweig}{1981}]{rosenzweig-ecology81}
Rosenzweig, M.~L. (1981).
\newblock {A} theory of habitat selection.
\newblock {\em {E}cology\/}~{\em 62}, 327--335.

\bibitem[\protect\citeauthoryear{Rueffler, Van~Dooren, and Metz}{Rueffler
  et~al.}{2006}]{rueffler_etal-amnat06}
Rueffler, C., T.~J. Van~Dooren, and J.~A. Metz (2006).
\newblock {T}he interplay between behavior and morphology in the evolutionary
  dynamics of resource specialization.
\newblock {\em {A}merican {N}aturalist\/}~{\em 169}, E34--E52.

\bibitem[\protect\citeauthoryear{Sauve, Fontaine, and Th{\'e}bault}{Sauve
  et~al.}{2016}]{sauve_etal-theorecol15}
Sauve, A. M.~C., C.~Fontaine, and E.~Th{\'e}bault (2016).
\newblock {S}tability of a diamond-shaped module with multiple interaction
  types.
\newblock {\em {T}heoretical {E}cology\/}~{\em 9}, 27--37.

\bibitem[\protect\citeauthoryear{Sheppard}{Sheppard}{2019}]{sheppard-bionv19}
Sheppard, C.~S. (2019).
\newblock {R}elative performance of co-occurring alien plant invaders depends
  on traits related to competitive ability more than niche differences.
\newblock {\em {B}iological {I}nvasions\/}~{\em 21}, 1101--1114.

\bibitem[\protect\citeauthoryear{Tilman}{Tilman}{1982}]{tilman1982}
Tilman, D. (1982).
\newblock {\em {R}esource {C}ompetition and {C}ommunity {S}tructure}.
\newblock Princeton: Princeton University Press.

\bibitem[\protect\citeauthoryear{Valdovinos, Moisset~de Espan{\'e}s, Flores,
  and Ramos-Jiliberto}{Valdovinos et~al.}{2013}]{valdovinos_etal-oikos13}
Valdovinos, F.~S., P.~Moisset~de Espan{\'e}s, J.~D. Flores, and
  R.~Ramos-Jiliberto (2013).
\newblock {A}daptive foraging allows the maintenance of biodiversity of
  pollination networks.
\newblock {\em {O}ikos\/}~{\em 122\/}(6), 907--917.

\bibitem[\protect\citeauthoryear{Vandermeer and Boucher}{Vandermeer and
  Boucher}{1978}]{vandermeer_boucher-jtb78}
Vandermeer, J. and D.~H. Boucher (1978).
\newblock {V}arieties of mutualistic interaction in population models.
\newblock {\em {J}ournal of {T}heoretical {B}iology\/}~{\em 74}, 549--558.

\bibitem[\protect\citeauthoryear{Waser and Real}{Waser and
  Real}{1979}]{waser_real-nature79}
Waser, N.~M. and L.~A. Real (1979).
\newblock {E}ffective mutualism between sequentially flowering plant species.
\newblock {\em {N}ature\/}~{\em 281}, 670--672.

\bibitem[\protect\citeauthoryear{Yan and Zhang}{Yan and
  Zhang}{2014}]{yan_zhang-rspb14}
Yan, C. and Z.~Zhang (2014).
\newblock {S}pecific non-monotonous interactions increase persistence of
  ecological networks.
\newblock {\em {P}roceedings of the {R}oyal {S}ociety {B}: {B}iological
  {S}ciences\/}~{\em 281}, 20132797.

\bibitem[\protect\citeauthoryear{Zhang, Yan, Krebs, and Stenseth}{Zhang
  et~al.}{2015}]{zhang_etal-ecomod15}
Zhang, Z., C.~Yan, C.~J. Krebs, and N.~C. Stenseth (2015).
\newblock {E}cological non-monotonicity and its effects on complexity and
  stability of populations, communities and ecosystems.
\newblock {\em {E}cological {M}odelling\/}~{\em 312}, 374--384.

\end{thebibliography}

\appendix
\renewcommand{\theequation}{A.\arabic{equation}}
\setcounter{equation}{0}
\renewcommand{\thefigure}{A.\arabic{figure}}
\setcounter{figure}{0}
\renewcommand{\thetable}{A.\arabic{table}}
\setcounter{table}{0}

\section{Plant population dynamics}

The switching line $e_{1}P_{1}=e_{2}P_{2}$ of the animal splits the positive quadrant
into

\[
\textrm{sector I}=\{(P_{1},P_{2})\;|\;e_{1}P_{1}>e_{2}P_{2},P_{1}\geq0,P_{2}\geq0\}
\]

\noindent and

\[
\textrm{sector II}=\{(P_{1},P_{2})\;|\;e_{1}P_{1}<e_{2}P_{2},P_{1}\geq0,P_{2}\geq0\}.
\]

In sector I animals interact with plant 1 only and plant population dynamics {[}system
(\ref{eq:plant_odes}) in the main text{]} are

\begin{equation}
\begin{aligned}\frac{dP_{1}}{dt} & =\left(r_{1}\left(1-\frac{P_{1}+c_{2}P_{2}}{K_{1}}\right)+sA\right)P_{1}\\
\frac{dP_{2}}{dt} & =\left(r_{2}\left(1-\frac{P_{2}+c_{1}P_{1}}{K_{2}}\right)\right)P_{2},
\end{aligned}
\label{eq:odeS1}
\end{equation}

\noindent whereas in sector II animals interact with plant 2 only and population
dynamics are

\begin{equation}
\begin{aligned}\frac{dP_{1}}{dt} & =\left(r_{1}\left(1-\frac{P_{1}+c_{2}P_{2}}{K_{1}}\right)\right)P_{1}\\
\frac{dP_{2}}{dt} & =\left(r_{2}\left(1-\frac{P_{2}+c_{1}P_{1}}{K_{2}}\right)+sA\right)P_{2}.
\end{aligned}
\label{eq:odeS2}
\end{equation}

Along the switching line $e_{1}P_{1}=e_{2}P_{2}$ animal strategy is not uniquely
defined and population dynamics satisfy

\begin{equation}
\begin{aligned}\frac{dP_{1}}{dt}= & r_{1}\left(1-\frac{P_{1}+c_{2}P_{2}}{K_{1}}\right)P_{1}+su_{1}P_{1}A\\
\frac{dP_{2}}{dt}= & r_{2}\left(1-\frac{P_{2}+c_{1}P_{1}}{K_{2}}\right)P_{2}+su_{2}P_{2}A\\
(u_{1},u_{2})\in & \{(v_{1},v_{2})\;|\;v_{1}+v_{2}=1,v_{1}\geq0,v_{2}\geq0\}.
\end{aligned}
\label{eq:ddi}
\end{equation}

\subsection{Plant dynamics in sectors I and II\label{sec:appdynsector}}

From (\ref{eq:odeS1}) and (\ref{eq:odeS2}), the isoclines of plant 1 in sectors
I and II are 
\begin{align}
P_{1}+c_{2}P_{2} & =K_{1}\left(1+\frac{sA}{r_{1}}\right)\label{eq:isoP1sectorI}\\
P_{1}+c_{2}P_{2} & =K_{1},\label{eq:isoP1sectorII}
\end{align}
respectively. We observe that plant 1 isocline exists in sector I iff $r_{1}+sA>0$.
For mutualists ($s=1$) this is always the case, but for exploiters this holds only
if $A<r_{1}$ which we assume now. The segment of plant 1 isocline in sector I given
in (\ref{eq:isoP1sectorI}) intersects the $P_{1}$ axis at $\mathbf{E_{1}}$ {[}given
by (\ref{eq:E1}) in the main text{]} and switching line (\ref{eq:A-switch}) at
\begin{equation}
\mathbf{b}=\left(\frac{e_{2}K_{1}(r_{1}+sA)}{r_{1}(e_{2}+c_{2}e_{1})}\,,\,\frac{e_{1}K_{1}(r_{1}+sA)}{r_{1}(e_{2}+c_{2}e_{1})}\right),\label{eq:b}
\end{equation}
and the segment of plant 1 isocline in sector II given in (\ref{eq:isoP1sectorII})
intersects the $P_{2}$ axis and the switching line at points 
\begin{align}
\mathbf{k_{1}} & =\left(0\,,\,\frac{K_{1}}{c_{2}}\right)\label{eq:k1}\\
\mathbf{a} & =\left(\frac{e_{2}K_{1}}{e_{2}+c_{2}e_{1}}\,,\,\frac{e_{1}K_{1}}{e_{2}+c_{2}e_{1}}\right),\label{eq:a}
\end{align}
respectively.

Similarly from (\ref{eq:odeS1}) and (\ref{eq:odeS2}), plant 2 isocline in sector
I is 
\begin{align}
P_{2}+c_{1}P_{1} & =K_{2}\label{eq:isoP2sectorI}
\end{align}
and in sector II 
\begin{align}
P_{2}+c_{1}P_{1} & =K_{2}\left(1+\frac{sA}{r_{2}}\right),\label{eq:isoP2sectorII}
\end{align}
respectively. Once again, plant 2 isocline exists in sector II iff $r_{2}+sA>0$.
Isocline (\ref{eq:isoP2sectorI}) intersects the $P_{1}$ axis and the switching
line at points 
\begin{align}
\mathbf{k_{2}} & =\left(\frac{K_{2}}{c_{1}}\,,\,0\right)\label{eq:k2}\\
\mathbf{q} & =\left(\frac{e_{2}K_{2}}{e_{1}+c_{1}e_{2}}\,,\,\frac{e_{1}K_{2}}{e_{1}+c_{1}e_{2}}\right),\label{eq:q}
\end{align}
respectively. Isocline (\ref{eq:isoP2sectorII}) intersects the $P_{2}$ axis at
$\mathbf{E_{2}}$ {[}given by (\ref{eq:E2}) in the main text{]} and the switching
line at 
\begin{equation}
\mathbf{p}=\left(\frac{e_{2}K_{2}(r_{2}+sA)}{r_{2}(e_{1}+c_{1}e_{2})}\,,\,\frac{e_{1}K_{2}(r_{2}+sA)}{r_{2}(e_{1}+c_{1}e_{2})}\right).\label{eq:p}
\end{equation}

Isoclines position in sector I is determined by position of $\mathbf{k_{2}}$ with
respect to $\mathbf{E_{1}}$ on the $P_{1}$ axis, and position of \textbf{b} with
respect to \textbf{q} along the switching line. The following statements apply in
this sector 
\begin{align}
\mathbf{k_{2}}>\mathbf{E_{1}} & \Longleftrightarrow\frac{K_{2}}{K_{1}}>c_{1}\left(1+\frac{sA}{r_{1}}\right)\equiv\gamma_{2}\label{eq:P2invades}\\
\mathbf{q}>\mathbf{b} & \Longleftrightarrow\frac{K_{2}}{K_{1}}>\left(\frac{e_{1}+c_{1}e_{2}}{e_{2}+c_{2}e_{1}}\right)\left(1+\frac{sA}{r_{1}}\right)\equiv\tau_{2}.\label{eq:P2trait}
\end{align}
If both conditions above are true, plant 2 isocline is above plant 1 isocline in
sector I and there is no interior equilibrium in this sector (e.g., Figure \ref{fig:GIclines}c,
sector I). If both conditions are false, then plant 1 isocline is above plant 2 isocline
in sector I (Figure \ref{fig:partable_antweak}c, sector I). If (\ref{eq:P2invades})
is true and (\ref{eq:P2trait}) false, isoclines intersect at point $\mathbf{E_{I}}$
{[}given by (\ref{eq:E.I}) in the main text{]}, and because plant 1 isocline is
steeper than plant 2 isocline $(\frac{1}{c_{2}}>c_{1})$ this equilibrium is stable
(e.g., Figure \ref{fig:partable_antweak}f, sector I). If (\ref{eq:P2invades}) is
false and (\ref{eq:P2trait}) true, isoclines intersect again but because plant 2
isocline is steeper than plant 1 isocline $(\frac{1}{c_{2}}<c_{1})$, $\mathbf{E_{I}}$
is unstable (e.g., Figure \ref{fig:partable_antstrong}d, sector I).

For sector II we compare $\mathbf{k_{1}}$ with $\mathbf{E_{2}}$ on the $P_{2}$
axis, and \textbf{a} with \textbf{p} along the switching line. We obtain 
\begin{align}
\mathbf{k_{1}}>\mathbf{E_{2}} & \Longleftrightarrow\frac{K_{1}}{K_{2}}>c_{2}\left(1+\frac{sA}{r_{2}}\right)\equiv\gamma_{1}\label{eq:P1invades}\\
\mathbf{a}>\mathbf{p} & \Longleftrightarrow\frac{K_{1}}{K_{2}}>\left(\frac{e_{2}+c_{2}e_{1}}{e_{1}+c_{1}e_{2}}\right)\left(1+\frac{sA}{r_{2}}\right)\equiv\tau_{1}.\label{eq:P1trait}
\end{align}
If both conditions above are true (e.g., Figure \ref{fig:GIclines}a,c) or both are
false (e.g., Figure \ref{fig:partable_antweak}g), there is no interior equilibrium
in sector II because the two plant isoclines do not intersect there. If (\ref{eq:P1invades})
is true and (\ref{eq:P1trait}) false, isoclines intersect at the point $\mathbf{E_{II}}$
{[}given by (\ref{eq:E.II}) in the main text{]}, and because plant 1 isocline is
steeper than plant 2 isocline $(\frac{1}{c_{2}}>c_{1})$ the equilibrium is stable
(e.g., Figure \ref{fig:GIclines}e, sector II). And if (\ref{eq:P1invades}) is false
and (\ref{eq:P1trait}) true, isoclines intersect and because plant 2 isocline is
steeper than plant 1 isocline $(\frac{1}{c_{2}}<c_{1})$, $\mathbf{E_{II}}$ is unstable
(e.g., Figure \ref{fig:partable_antstrong}b, sector II).

\subsection{Plant population dynamics along the switching line\label{sec:appsliding}}

\noindent Here we are interested in plant population dynamics at the switching line.
Let $\mathbf{n}=(e_{1},-e_{2})$ be a perpendicular vector to the switching line
$e_{1}P_{1}=e_{2}P_{2}$ and let us denote the right hand sides of (\ref{eq:odeS1})
and (\ref{eq:odeS2}) by $\mathbf{f^{I}}$ and $\mathbf{f^{II}}$, respectively.
The dynamics close to the switching line depend on the following scalar products
\begin{equation}
\begin{aligned}\langle\mathbf{n,f^{I}}\rangle & =e_{1}P_{1}\left\{ (r_{1}+sA)-r_{2}+P_{1}\frac{K_{1}r_{2}(e_{1}+c_{1}e_{2})-K_{2}r_{1}(e_{2}+c_{2}e_{1})}{e_{2}K_{1}K_{2}}\right\} \\
\langle\mathbf{n,f^{II}}\rangle & =e_{1}P_{1}\left\{ r_{1}-(r_{2}+sA)+P_{1}\frac{K_{1}r_{2}(e_{1}+c_{1}e_{2})-K_{2}r_{1}(e_{2}+c_{2}e_{1})}{e_{2}K_{1}K_{2}}\right\} .
\end{aligned}
\label{ineq}
\end{equation}

There are four possibilities \citep{filippov1988,colombo_krivan-jmamb93}: 
\begin{enumerate}
\item If $\langle\mathbf{n,f^{I}}\rangle<0$ and $\langle\mathbf{n,f^{II}}\rangle<0$ trajectories
are crossing the switching line in direction from sector I to sector II. 
\item If $\langle\mathbf{n,f^{I}}\rangle>0$ and $\langle\mathbf{n,f^{II}}\rangle>0$ trajectories
are crossing the switching line in direction from sector II to sector I. 
\item If $\langle\mathbf{n,f^{I}}\rangle<0$ and $\langle\mathbf{n,f^{II}}\rangle>0$ trajectories
do not cross the switching line and they have to stay for some positive time on the
switching line. This is called the sliding regime. 
\item If $\langle\mathbf{n,f^{I}}\rangle>0$ and $\langle\mathbf{n,f^{II}}\rangle<0$ trajectories
that start at such points are not uniquely defined. They can move along the switching
line for some time and then leave the line either to sector I or to sector II. This
is called the repelling regime. 
\end{enumerate}
We observe that 
\[
\langle\mathbf{n,f^{I}}\rangle=\langle\mathbf{n,f^{II}}\rangle+2se_{1}P_{1}A.
\]
Thus, when $s=1$, $\langle\mathbf{n,f^{II}}\rangle>0$ implies $\langle\mathbf{n,f^{I}}\rangle>0$
which excludes the sliding regime. Similarly, when $s=-1$, $\langle\mathbf{n,f^{II}}\rangle<0$
implies $\langle\mathbf{n,f^{I}}\rangle<0$ which excludes the repelling regime.

To analyze all possible situations under which sliding or repelling regime occurs,
using (\ref{eq:b}), (\ref{eq:a}), (\ref{eq:q}), and (\ref{eq:p}) we rewrite (\ref{ineq})
as 
\begin{equation}
\begin{aligned}\langle\mathbf{n,f^{I}}\rangle & =\frac{e_{1}P_{1}\left\{ K_{1}r_{2}(e_{1}+c_{1}e_{2})\left[P_{1}-q_{1}\right]-K_{2}r_{1}(e_{2}+c_{2}e_{1})\left[P_{1}-b_{1}\right]\right\} }{e_{2}K_{1}K_{2}}\\
\langle\mathbf{n,f^{II}}\rangle & =\frac{e_{1}P_{1}\left\{ K_{1}r_{2}(e_{1}+c_{1}e_{2})\left[P_{1}-p_{1}\right]-K_{2}r_{1}(e_{2}+c_{2}e_{1})\left[P_{1}-a_{1}\right]\right\} }{e_{2}K_{1}K_{2}}.
\end{aligned}
\label{ineq2}
\end{equation}
For exploiters ($s=-1$) $\mathbf{b}<\mathbf{a}$ and $\mathbf{p}<\mathbf{q}$ so
that there are four possibilities for isoclines overlap at the switching line. All
these possibilities together with the overlap segment of the two generalized isoclines
are listed in Table \ref{tab:switchline_dynamics}. Moreover, scalar products given
in (\ref{ineq2}) show that in the overlap segment plant dynamics are in the sliding
regime.

Similarly, for mutualists ($s=1$) $\mathbf{b}>\mathbf{a}$ and $\mathbf{p}>\mathbf{q}$
and again there are four possibilities where the two isoclines overlap at the switching
line (Table \ref{tab:switchline_dynamics}). However, in this case, the overlap segment
repels trajectories. 
\begin{center}
\begin{table}
\begin{centering}
\begin{tabular}{ccccccl}
 & Cases  & $\langle\mathbf{n,f^{I}}\rangle$  & $\langle\mathbf{n,f^{II}}\rangle$  & overlap  & $\mathbf{E_{S}}$  & dynamics at the \tabularnewline
 &  &  &  & segment  &  & overlap segment \tabularnewline
\hline 
$s=-1$  & $p_{1}<b_{1}<a_{1}<q_{1}$  & <0  & >0  & $\mathbf{ba}$  & Yes  & sliding regime \tabularnewline
 & $b_{1}<p_{1}<q_{1}<a_{1}$  & <0  & >0  & $\mathbf{pq}$  & Yes  & sliding regime \tabularnewline
 & $p_{1}<b_{1}<q_{1}<a_{1}$  & <0  & >0  & $\mathbf{bq}$  & Yes  & sliding regime \tabularnewline
 & $b_{1}<p_{1}<a_{1}<q_{1}$  & <0  & >0  & $\mathbf{pa}$  & Yes  & sliding regime \tabularnewline
 & $p_{1}<q_{1}<b_{1}<a_{1}$  & >0  & >0  & no overlap  & No  & crossing from sector II to I\tabularnewline
 & $b_{1}<a_{1}<p_{1}<q_{1}$  & <0  & <0  & no overlap  & No  & crossing from sector I to II\tabularnewline
\hline 
$s=1$  & $q_{1}<a_{1}<b_{1}<p_{1}$  & >0  & <0  & $\mathbf{ab}$  & Yes  & repelling regime \tabularnewline
 & $a_{1}<q_{1}<p_{1}<b_{1}$  & >0  & <0  & $\mathbf{qp}$  & Yes  & repelling regime \tabularnewline
 & $q_{1}<a_{1}<p_{1}<b_{1}$  & >0  & <0  & $\mathbf{ap}$  & Yes  & repelling regime \tabularnewline
 & $a_{1}<q_{1}<b_{1}<p_{1}$  & >0  & <0  & $\mathbf{qb}$  & Yes  & repelling regime \tabularnewline
 & $q_{1}<p_{1}<a_{1}<b_{1}$  & >0  & >0  & no overlap  & No  & crossing from sector II to I\tabularnewline
 & $a_{1}<b_{1}<q_{1}<p_{1}$  & <0  & <0  & no overlap  & No  & crossing from sector I to II\tabularnewline
\hline 
\end{tabular}
\par\end{centering}
\caption{\label{tab:switchline_dynamics} List of all possible overlaps of generalized isoclines
along the switching line.}
\end{table}
\par\end{center}

\subsubsection{Equilibrium $\mathbf{E_{S}}$}

Now we look for equilibria of model (\ref{eq:plant_odes}) and (\ref{eq:pref_step})
in the switching line. Every non-trivial equilibrium there must satisfy 
\[
\begin{aligned}e_{1}P_{1}= & e_{2}P_{2}\\
0= & r_{1}\left(1-\frac{P_{1}+c_{2}P_{2}}{K_{1}}\right)P_{1}+su_{1}P_{1}A\\
0= & r_{2}\left(1-\frac{P_{2}+c_{1}P_{1}}{K_{2}}\right)P_{2}+s(1-u_{1})P_{2}A.
\end{aligned}
\]
These equations have a single non-trivial solution that gives equilibrium $\mathbf{E_{S}}$
given in (\ref{eq:E.S}) and the corresponding preference for plant 1, $\bar{u}_{1}$,
given in (\ref{eq:U.S}). For $\mathbf{E_{S}}$ to be feasible, $\bar{u}_{1}$ must
be between 0 and 1. This happens iff either (\ref{eq:cond_attract}) or (\ref{eq:cond_repel})
holds. Using (\ref{eq:b}) and (\ref{eq:q}), plant 1 population equilibrium given
in (\ref{eq:E.S}) can be written as a convex combination of points $b_{1}$ and
$q_{1}$

\[
\bar{P}_{1}=\left[\frac{K_{2}r_{1}(e_{2}+c_{2}e_{1})}{K_{1}r_{2}(e_{1}+c_{1}e_{2})+K_{2}r_{1}(e_{2}+c_{2}e_{1})}\right]b_{1}+\left[\frac{K_{1}r_{2}(e_{1}+c_{1}e_{2})}{K_{1}r_{2}(e_{1}+c_{1}e_{2})+K_{2}r_{1}(e_{2}+c_{2}e_{1})}\right]q_{1},
\]
which shows that $b_{1}<\bar{P}_{1}<q_{1}.$

\noindent Similarly, using (\ref{eq:a}) and (\ref{eq:p}), plant 1 population equilibrium
becomes

\[
\bar{P}_{1}=\left[\frac{K_{2}r_{1}(e_{2}+c_{2}e_{1})}{K_{1}r_{2}(e_{1}+c_{1}e_{2})+K_{2}r_{1}(e_{2}+c_{2}e_{1})}\right]a_{1}+\left[\frac{K_{1}r_{2}(e_{1}+c_{1}e_{2})}{K_{1}r_{2}(e_{1}+c_{1}e_{2})+K_{2}r_{1}(e_{2}+c_{2}e_{1})}\right]p_{1}
\]

\noindent which shows that $p_{1}<\bar{P}_{1}<a_{1}$. It follows from Table \ref{tab:switchline_dynamics}
that equilibrium $\mathbf{E_{S}}$ is in the sliding regime where the plant generalized
isoclines overlap. Now we study stability of $\mathbf{E_{S}}$.

First we consider the exploitation case where $s=-1.$ Table \ref{tab:switchline_dynamics}
shows that at points where the generalized isoclines overlap, trajectories are driven
toward the switching line from both sectors. In this case trajectories cannot cross
the switching line inside the isoclines overlap segment. Thus, once a trajectory
reaches the overlap segment, it must move along it, i.e., $e_{1}P_{1}(t)=e_{2}P_{2}(t)$.
This means that when the trajectory moves along the overlap segment, preferences
for plants $(u_{1},u_{2})$ must satisfy $e_{1}P_{1}'(t)=e_{2}P_{2}'(t),$ i.e.,

\[
e_{1}\left[r_{1}\left(1-\frac{P_{1}+c_{2}P_{2}}{K_{1}}\right)+su_{1}A\right]=e_{2}\left[r_{2}\left(1-\frac{P_{2}+c_{1}P_{1}}{K_{2}}+s(1-u_{1})A\right)\right],
\]

\noindent where we used the fact that $e_{1}P_{1}(t)=e_{2}P_{2}(t)$. The corresponding
preference for plant plant 1 along the trajectory is

\[
u_{1}=\frac{e_{2}K_{2}(sAe_{1}K_{1}+c_{2}e_{1}P_{1}r_{1}+e_{2}r_{1}(P_{1}-K_{1}))-e_{1}K_{1}r_{2}(c_{1}e_{2}P_{1}+e_{1}P_{1}-e_{2}K_{2})}{sAe_{2}K_{1}K_{2}(e_{1}+e_{2})}.
\]

With this preference for plant 1, plant population dynamics in the sliding regime
are described by the logistic equation

\begin{equation}
\frac{dP_{1}}{dt}=\frac{e_{1}(r_{1}+r_{2}+sA)}{e_{1}+e_{2}}\left[1-\left(\frac{K_{1}r_{2}(e_{1}+c_{1}e_{2})+K_{2}r_{1}(e_{2}+c_{2}e_{1})}{e_{2}K_{1}K_{2}(r_{1}+r_{2}+sA)}\right)P_{1}\right]P_{1},\label{eq:logistic_switching}
\end{equation}

\noindent with equilibrium $\bar{P}_{1}$ corresponding to $\mathbf{E_{S}}=(\bar{P}_{1},e_{1}/e_{2}\bar{P}_{1}).$
This shows that equilibrium $\mathbf{E_{S}}$ is locally stable, because trajectories
close to this equilibrium are attracted from both sector I and II toward the switching
line (Table \ref{tab:switchline_dynamics}) and they converge along the switching
line to the equilibrium.

Second, we consider stability of $\mathbf{E_{S}}$ for mutualisms when $s=1$. Table
\ref{tab:switchline_dynamics} shows that the overlap segment of the two isoclines
repels nearby trajectories, equilibrium $\mathbf{E_{S}}$ is unstable. Moreover,
trajectories that start at the overlap of the two plant generalized isoclines are
not uniquely defined, because they can leave this segment of the switching line either
to sector I, or to sector II.

\section{Effect of parameters on equilibria\label{sec:appareffect}}

Using (\ref{eq:isoclines}) and (\ref{eq:Hi}) for sector I, equilibrium densities
at $\mathbf{E_{I}}$ (\ref{eq:E.I}) take the form 
\[
\hat{P}_{1}=\frac{H_{1}-c_{2}K_{2}}{1-c_{1}c_{2}}\,,\,\hat{P}_{2}=\frac{K_{2}-c_{1}H_{1}}{1-c_{1}c_{2}},
\]
where $H_{1}=K_{1}\left(1+\frac{sA}{r_{1}}\right)$. Thus, $\partial\hat{P}_{i}/\partial r_{2},\partial\hat{P}_{i}/\partial e_{1},\partial\hat{P}_{i}/\partial e_{2}\,(i=1,2)$
are all zero, and 
\begin{align*}
\frac{\partial\hat{P}_{1}}{\partial r_{1}} & =\frac{-sK_{1}A}{r_{1}^{2}(1-c_{1}c_{2})}, & \frac{\partial\hat{P}_{2}}{\partial r_{1}} & =\frac{sc_{1}K_{1}A}{r_{1}^{2}(1-c_{1}c_{2})}, & \frac{\partial\hat{P}_{1}}{\partial A} & =\frac{sK_{1}}{r_{1}(1-c_{1}c_{2})}, & \frac{\partial\hat{P}_{2}}{\partial A} & =\frac{-sK_{1}c_{1}}{r_{1}(1-c_{1}c_{2})},\\
\frac{\partial\hat{P}_{1}}{\partial K_{1}} & =\frac{1}{1-c_{1}c_{2}}\left(1+\frac{sA}{r_{1}}\right), & \frac{\partial\hat{P}_{2}}{\partial K_{1}} & =\frac{-c_{1}}{1-c_{1}c_{2}}\left(1+\frac{sA}{r_{1}}\right), & \frac{\partial\hat{P}_{1}}{\partial K_{2}} & =\frac{-c_{2}}{1-c_{1}c_{2}}, & \frac{\partial\hat{P}_{2}}{\partial K_{2}} & =\frac{1}{1-c_{1}c_{2}},\\
\frac{\partial\hat{P}_{1}}{\partial c_{1}} & =\frac{c_{2}\hat{P}_{1}}{1-c_{1}c_{2}}, & \frac{\partial\hat{P}_{2}}{\partial c_{1}} & =\frac{-\hat{P}_{1}}{1-c_{1}c_{2}}, & \frac{\partial\hat{P}_{1}}{\partial c_{2}} & =\frac{-\hat{P}_{2}}{1-c_{1}c_{2}}, & \frac{\partial\hat{P}_{2}}{\partial c_{2}} & =\frac{c_{1}\hat{P}_{2}}{1-c_{1}c_{2}}.
\end{align*}
We remark that because $r_{1}+sA>0$ is required for $\mathbf{E_{I}}$ to be feasible,
the sign of $\frac{\partial\hat{P}_{1}}{\partial K_{1}}$ and $\frac{\partial\hat{P}_{2}}{\partial K_{1}}$
is independent of $1+\frac{sA}{r_{1}}$. Parameter effects on $\mathbf{E_{II}}$
are obtained analogously.

At equilibrium $\mathbf{E_{S}}$ (\ref{eq:E.S}) plant densities take the form 
\[
\bar{P}_{1}=e_{2}G\,,\,\bar{P}_{2}=e_{1}G,
\]
where $G=\dfrac{K_{1}K_{2}(r_{1}+r_{2}+sA)}{K_{1}r_{2}(e_{1}+c_{1}e_{2})+K_{2}r_{1}(e_{2}+c_{2}e_{1})}$.
This quantity varies with parameters as 
\begin{align*}
\frac{\partial G}{\partial r_{1}} & =\left(\frac{K_{1}}{K_{2}}-\tau_{1}\right)\left\{ \dfrac{r_{2}(e_{1}+c_{1}e_{2})K_{1}K_{2}^{2}}{\left[K_{1}r_{2}(e_{1}+c_{1}e_{2})+K_{2}r_{1}(e_{2}+c_{2}e_{1})\right]^{2}}\right\} \\
\frac{\partial G}{\partial K_{1}} & =\left\{ \dfrac{r_{1}(r_{1}+r_{2}+sA)(e_{2}+c_{2}e_{1})K_{2}^{2}}{\left[K_{1}r_{2}(e_{1}+c_{1}e_{2})+K_{2}r_{1}(e_{2}+c_{2}e_{1})\right]^{2}}\right\} \\
\frac{\partial G}{\partial c_{1}} & =-\left\{ \dfrac{e_{2}r_{2}(r_{1}+r_{2}+sA)K_{1}^{2}K_{2}}{\left[K_{1}r_{2}(e_{1}+c_{1}e_{2})+K_{2}r_{1}(e_{2}+c_{2}e_{1})\right]^{2}}\right\} \\
\frac{\partial G}{\partial A} & =s\left\{ \dfrac{K_{1}K_{2}}{K_{1}r_{2}(e_{1}+c_{1}e_{2})+K_{2}r_{1}(e_{2}+c_{2}e_{1})}\right\} \\
\frac{\partial G}{\partial e_{1}} & =-\left\{ \dfrac{K_{1}K_{2}(r_{1}+r_{2}+sA)(K_{1}r_{2}+K_{2}r_{1}c_{2})}{\left[K_{1}r_{2}(e_{1}+c_{1}e_{2})+K_{2}r_{1}(e_{2}+c_{2}e_{1})\right]^{2}}\right\} ,
\end{align*}
where the quantities between curly braces are positive (because feasibility of $\mathbf{E_{S}}$
requires $r_{1}+r_{2}+sA>0$). Thus $\frac{\partial G}{\partial K_{1}}>0,\frac{\partial G}{\partial c_{1}}<0,$
and $\frac{\partial G}{\partial e_{1}}<0$. Moreover, $\frac{\partial G}{\partial A}<0$
under exploitation $(s=-1)$ and $\frac{\partial G}{\partial A}>0$ under mutualism
$(s=1)$. Under exploitation $\frac{\partial G}{\partial r_{1}}>0$ because $\mathbf{E_{S}}$
is feasible iff both plants are \emph{above }their attraction thresholds (i.e., $K_{1}/K_{2}>\tau_{1}$
and $K_{2}/K_{1}>\tau_{2}$). Conversely, $\frac{\partial G}{\partial r_{1}}<0$
under mutualism. Since $\bar{P}_{i}=e_{j}G$ where $i,j=1,2$ but $i\neq j$, we
can conclude 
\begin{align*}
\frac{\partial\bar{P}_{i}}{\partial K_{1}} & >0, & \frac{\partial\bar{P}_{i}}{\partial c_{1}} & <0, & \frac{\partial\bar{P}_{i}}{\partial r_{1}} & \begin{cases}
>0 & \textrm{exploitation}\\
<0 & \textrm{mutualism},
\end{cases} & \frac{\partial\bar{P}_{i}}{\partial A} & \begin{cases}
<0 & \textrm{exploitation}\\
>0 & \textrm{mutualism},
\end{cases},
\end{align*}
i.e., both plant densities change in the same direction (i.e., $\partial\bar{P}_{1}/\partial\bar{P}_{2}>0$)
when $r_{1},K_{1},c_{1},A$ change. Now when $e_{1}$ varies we have $\frac{\partial\bar{P}_{1}}{\partial e_{1}}=e_{2}\frac{\partial G}{\partial e_{1}}<0$,
but 
\[
\frac{\partial\bar{P}_{2}}{\partial e_{1}}=G+e_{1}\frac{\partial G}{\partial r_{1}}=\dfrac{e_{2}K_{1}K_{2}(K_{2}r_{1}+c_{1}K_{1}r_{2})(r_{1}+r_{2}+sA)}{\left[K_{1}r_{2}(e_{1}+c_{1}e_{2})+K_{2}r_{1}(e_{2}+c_{2}e_{1})\right]^{2}}
\]
which is positive. \emph{Mutatis mutandis} $\frac{\partial\bar{P}_{1}}{\partial e_{2}}>0$
and $\frac{\partial\bar{P}_{2}}{\partial e_{2}}<0$. Thus, when $e_{1}$ or $e_{2}$
change, plant densities change in opposite directions (i.e., $\partial\bar{P}_{1}/\partial\bar{P}_{2}<0$).

Finally the derivatives of generalist preference $\bar{u}_{1}$ (\ref{eq:U.S}) at
$\mathbf{E_{S}}$ are 
\begin{align*}
\frac{\partial\bar{u}_{1}}{\partial r_{1}} & =\left\{ \frac{r_{2}(e_{2}+c_{2}e_{1})K_{1}\,\bar{u}_{1}}{r_{1}\left[K_{1}r_{2}(e_{1}+c_{1}e_{2})+K_{2}r_{1}(e_{2}+c_{2}e_{1})\right]}\right\} \\
\frac{\partial\bar{u}_{1}}{\partial K_{1}} & =-s\left\{ \dfrac{r_{1}r_{2}K_{2}(e_{1}+c_{1}e_{2})(e_{2}+c_{2}e_{1})(r_{1}+r_{2}+sA)}{A\left[K_{1}r_{2}(e_{1}+c_{1}e_{2})+K_{2}r_{1}(e_{2}+c_{2}e_{1})\right]^{2}}\right\} \\
\frac{\partial\bar{u}_{1}}{\partial c_{1}} & =-s\left\{ \dfrac{r_{1}r_{2}e_{2}K_{1}K_{2}(e_{2}+c_{2}e_{1})(r_{1}+r_{2}+sA)}{A\left[K_{1}r_{2}(e_{1}+c_{1}e_{2})+K_{2}r_{1}(e_{2}+c_{2}e_{1})\right]^{2}}\right\} \\
\frac{\partial\bar{u}_{1}}{\partial e_{1}} & =s(c_{1}c_{2}-1)\left\{ \dfrac{e_{2}r_{1}r_{2}K_{1}K_{2}(r_{1}+r_{2}+sA)}{A\left[K_{1}r_{2}(e_{1}+c_{1}e_{2})+K_{2}r_{1}(e_{2}+c_{2}e_{1})\right]^{2}}\right\} \\
\frac{\partial\bar{u}_{1}}{\partial A} & =s\left(\frac{K_{1}}{K_{2}}-\frac{e_{2}+c_{2}e_{1}}{e_{1}+c_{1}e_{2}}\right)\left\{ \dfrac{r_{1}r_{2}K_{2}(e_{1}+c_{1}e_{2})}{A^{2}\left[K_{1}r_{2}(e_{1}+c_{1}e_{2})+K_{2}r_{1}(e_{2}+c_{2}e_{1})\right]}\right\} ,
\end{align*}
where the quantities between curly braces are positive (because feasibility of $\mathbf{E_{S}}$
requires $r_{1}+r_{2}+sA>0$). Thus $\frac{\partial\bar{u}_{1}}{\partial r_{1}}>0$
trivially. Under exploitation $(s=-1)$, $\frac{\partial\bar{u}_{1}}{\partial K_{1}}>0$,
and $\frac{\partial\bar{u}_{1}}{\partial c_{1}}>0$. And under mutualism $(s=1)$,
$\frac{\partial\bar{u}_{1}}{\partial K_{1}}<0$ and $\frac{\partial\bar{u}_{1}}{\partial c_{1}}<0$.
The sign of $\frac{\partial\bar{u}_{1}}{\partial e_{1}}$ depends on interaction
type and strength of competition as follows 
\[
\frac{\partial\bar{u}_{1}}{\partial e_{1}}\begin{cases}
>0 & \textrm{for exploitation \& }c_{1}c_{2}<1\\
<0 & \textrm{for exploitation \& }c_{1}c_{2}>1\\
<0 & \textrm{for mutualism \& }c_{1}c_{2}<1\\
>0 & \textrm{for mutualism \& }c_{1}c_{2}>1
\end{cases}
\]
and the sign of $\partial\bar{u}_{1}/\partial A$ depends on interaction type and
the sign of the $\frac{K_{1}}{K_{2}}-\frac{e_{2}+c_{2}e_{1}}{e_{1}+c_{1}e_{2}}$.
Since $u_{1}$ and $u_{2}$ vary in opposite directions, the derivatives of $\bar{u}_{1}$
with respect to $r_{2},K_{2},c_{2},e_{2}$ are of opposite signs compared with the
corresponding derivatives with respect $r_{1},K_{1},c_{1},e_{1}$ above.

\section{Classification of equilibria\label{sec:appclassification}}

Table \ref{tab:stable_equilibria} summarizes our previous analyzes given in Appendices
\ref{sec:appdynsector} and \ref{sec:appsliding}, and lists all feasible (i.e.,
non-negative) stable equilibria for system (\ref{eq:plant_odes}) under exploitation
$(s=-1)$ or mutualism $(s=1)$, and weak $(c_{1}c_{2}<1)$ or strong $(c_{1}c_{2}>1)$
competition. Empty $(\emptyset)$ cells indicate that no parameter combination satisfies
row or column conditions. Cells with only one equilibrium indicate that this equilibrium
is globally stable. Cells with multiple equilibria indicate that these equilibria
are locally stable. There are 56 non-empty cells in Table \ref{tab:stable_equilibria},
each of them corresponding to a unique isocline configuration. The configurations
shown in Figures \ref{fig:partable_antweak}\textendash \ref{fig:partable_mutstrong}
are indicated by figure number and panel. Out of these 56 configurations, there are
11 possible combinations (i.e., $\mathbf{E_{1}}$, $\mathbf{E_{2}}$, $\mathbf{E_{I}}$,
$\mathbf{E_{II}}$, $\mathbf{E_{S}}$, $\{\mathbf{E_{I}},\mathbf{E_{II}}\}$, $\{\mathbf{E_{S}},\mathbf{E_{1}}\}$,
$\{\mathbf{E_{S}},\mathbf{E_{2}}\}$, $\{\mathbf{E_{1}},\mathbf{E_{II}}\}$, $\{\mathbf{E_{2}},\mathbf{E_{I}}\}$,
$\{\mathbf{E_{S}},\mathbf{E_{1}},\mathbf{E_{2}}\}$) with respect to stable equilibria.

Equilibria $\mathbf{E_{1},E_{2}}$ given in (\ref{eq:E1}), (\ref{eq:E2}) are boundary
(i.e., monoculture) equilibria for plant 1 and 2, respectively; $\mathbf{E_{I},E_{II},E_{S}}$
given in (\ref{eq:E.I}), (\ref{eq:E.II}), and (\ref{eq:E.S}) are interior equilibria
in sector I (where $u_{1}=1$), sector II (where $u_{1}=0$), and the switching line
(where $u_{1}=\bar{u}_{1}$ is given by (\ref{eq:U.S})), respectively. Cases are
classified with respect to position of $\mathbf{k_{1}}$ given in (\ref{eq:k1})
and $\mathbf{E_{1}}$ on $P_{1}$ axes, $\mathbf{k_{2}}$ given in (\ref{eq:k2})
and $\mathbf{E_{2}}$ on $P_{2}$ axes, and points \textbf{a}, \textbf{b}, \textbf{p},
\textbf{q} given in (\ref{eq:a}), (\ref{eq:b}), (\ref{eq:p}), (\ref{eq:q}) along
the switching line. For mutualisms ($s=1$), $\mathbf{q}<\mathbf{p}$ and $\mathbf{a}<\mathbf{b}$
while for exploitation ($s=-1$), $\mathbf{p}<\mathbf{q}$ and $\mathbf{b}<\mathbf{a}.$
We remark that for exploitation when $A>r_{1}$ ($A>r_{2}$), point $\mathbf{b}$
($\mathbf{p}$) is in the third quadrant and boundary equilibrium $\mathbf{E_{1}}$
($\mathbf{E_{2}}$) is not feasible. Table \ref{tab:stable_equilibria} considers
all generic cases excluding those cases where one or more inequalities between points
and parameters are replaced by equalities. 
\begin{center}
\begin{sidewaystable}
\begin{centering}
\begin{tabular}{|c|c|c||c||c||c||c||c||c||c||c||c||c||c||c||c||c|>{\centering}p{3.5cm}|>{\centering}p{3.5cm}|>{\centering}p{3.5cm}|>{\centering}p{4cm}|}
\hline 
\multicolumn{2}{|c|}{Interacting} & \multicolumn{15}{c|}{Position of isoclines intersections} & \multicolumn{4}{c|}{Position of isoclines intersections along $P_{1}$ and $P_{2}$ axes}\tabularnewline
\cline{18-21} 
\multicolumn{2}{|c|}{conditions} & \multicolumn{15}{c|}{along the switching line} & $\mathbf{k_{1}>E_{2}},\mathbf{k_{2}>E_{1}}$ & $\mathbf{k_{1}>E_{2}},\mathbf{k_{2}<E_{1}}$ & $\mathbf{k_{1}<E_{2}},\mathbf{k_{2}>E_{1}}$ & $\mathbf{k_{1}<E_{2}},\mathbf{k_{2}<E_{1}}$\tabularnewline
\hline 
\multirow{12}{*}{$s=-1$} & \multirow{6}{*}{$c_{1}c_{2}<1$} & \multicolumn{15}{c|}{$\mathbf{p<q<b<a}$ } & $\mathbf{E_{I}}$(Fig. \ref{fig:partable_antweak}f) & $\mathbf{E_{1}}$(Fig. \ref{fig:partable_antweak}c) & $\emptyset$ & $\emptyset$\tabularnewline
 &  & \multicolumn{15}{c|}{$\mathbf{b<a<p<q}$ } & $\mathbf{E_{II}}$(Fig. \ref{fig:partable_antweak}h) & $\emptyset$ & $\mathbf{E_{2}}$(Fig. \ref{fig:partable_antweak}g) & $\emptyset$\tabularnewline
 &  & \multicolumn{15}{c|}{$\mathbf{p<b<a<q}$ } & $\mathbf{E_{S}}$ & $\emptyset$ & $\emptyset$ & $\emptyset$\tabularnewline
 &  & \multicolumn{15}{c|}{$\mathbf{b<p<q<a}$ } & $\mathbf{E_{S}}$ & $\emptyset$ & $\emptyset$ & $\emptyset$\tabularnewline
 &  & \multicolumn{15}{c|}{$\mathbf{p<b<q<a}$ } & $\mathbf{E_{S}}$ & $\emptyset$ & $\emptyset$ & $\emptyset$\tabularnewline
 &  & \multicolumn{15}{c|}{$\mathbf{b<p<a<q}$ } & $\mathbf{E_{S}}$(Fig. \ref{fig:partable_antweak}i) & $\emptyset$ & $\emptyset$ & $\emptyset$\tabularnewline
\cline{2-21} 
 & \multirow{6}{*}{$c_{1}c_{2}>1$} & \multicolumn{15}{c|}{$\mathbf{p<q<b<a}$ } & $\emptyset$ & $\mathbf{E_{1}}$(Fig. \ref{fig:partable_antstrong}c) & $\emptyset$ & $\mathbf{\{E_{1},E_{2}\}}$(Fig. \ref{fig:partable_antstrong}b)\tabularnewline
 &  & \multicolumn{15}{c|}{$\mathbf{b<a<p<q}$ } & $\emptyset$ & $\emptyset$ & $\mathbf{E_{2}}$(Fig. \ref{fig:partable_antstrong}g) & $\mathbf{\{E_{1},E_{2}\}}$(Fig. \ref{fig:partable_antstrong}d)\tabularnewline
 &  & \multicolumn{15}{c|}{$\mathbf{p<b<a<q}$ } & $\mathbf{E_{S}}$ & $\mathbf{\{E_{S},E_{1}\}}$(Fig. \ref{fig:partable_antstrong}f) & $\mathbf{\{E_{S},E_{2}\}}$ & $\mathbf{\{E_{S},E_{1},E_{2}\}}$(Fig. \ref{fig:partable_antstrong}e)\tabularnewline
 &  & \multicolumn{15}{c|}{$\mathbf{b<p<q<a}$ } & $\mathbf{E_{S}}$ (Fig. \ref{fig:partable_antstrong}i) & $\mathbf{\{E_{S},E_{1}\}}$ & $\mathbf{\{E_{S},E_{2}\}}$ & $\mathbf{\{E_{S},E_{1},E_{2}\}}$\tabularnewline
 &  & \multicolumn{15}{c|}{$\mathbf{p<b<q<a}$ } & $\mathbf{E_{S}}$ & $\mathbf{\{E_{S},E_{1}\}}$ & $\mathbf{\{E_{S},E_{2}\}}$ & $\mathbf{\{E_{S},E_{1},E_{2}\}}$\tabularnewline
 &  & \multicolumn{15}{c|}{$\mathbf{b<p<a<q}$ } & $\mathbf{E_{S}}$ & $\mathbf{\{E_{S},E_{1}\}}$ & $\mathbf{\{E_{S},E_{2}\}}$(Fig. \ref{fig:partable_antstrong}h) & $\mathbf{\{E_{S},E_{1},E_{2}\}}$\tabularnewline
\hline 
\multirow{12}{*}{$s=1$ } & \multirow{6}{*}{$c_{1}c_{2}<1$} & \multicolumn{15}{c|}{$\mathbf{q<p<a<b}$ } & $\mathbf{E_{I}}$(Fig. \ref{fig:partable_mutweak}f) & $\mathbf{E_{1}}$(Fig. \ref{fig:partable_mutweak}c) & $\emptyset$ & $\emptyset$\tabularnewline
 &  & \multicolumn{15}{c|}{$\mathbf{a<b<q<p}$ } & $\mathbf{E_{II}}$(Fig. \ref{fig:partable_mutweak}h) & $\emptyset$ & $\mathbf{E_{2}}$(Fig. \ref{fig:partable_mutweak}g) & $\emptyset$\tabularnewline
 &  & \multicolumn{15}{c|}{$\mathbf{q<a<b<p}$ } & $\mathbf{\{E_{I},E_{II}\}}$ & $\mathbf{\{E_{II},E_{1}\}}$ & $\mathbf{\{E_{I},E_{2}\}}$ & $\mathbf{\{E_{1},E_{2}\}}$(Fig. \ref{fig:partable_mutweak}a)\tabularnewline
 &  & \multicolumn{15}{c|}{$\mathbf{a<q<p<b}$ } & $\mathbf{\{E_{I},E_{II}\}}$(Fig. \ref{fig:partable_mutweak}e) & $\mathbf{\{E_{II},E_{1}\}}$ & $\mathbf{\{E_{I},E_{2}\}}$ & $\mathbf{\{E_{1},E_{2}\}}$\tabularnewline
 &  & \multicolumn{15}{c|}{$\mathbf{q<a<p<b}$ } & $\mathbf{\{E_{I},E_{II}\}}$ & $\mathbf{\{E_{II},E_{1}\}}$(Fig. \ref{fig:partable_mutweak}b) & $\mathbf{\{E_{I},E_{2}\}}$ & $\mathbf{\{E_{1},E_{2}\}}$\tabularnewline
 &  & \multicolumn{15}{c|}{$\mathbf{a<q<b<p}$ } & $\mathbf{\{E_{I},E_{II}\}}$ & $\mathbf{\{E_{II},E_{1}\}}$ & $\mathbf{\{E_{I},E_{2}\}}$(Fig. \ref{fig:partable_mutweak}d) & $\mathbf{\{E_{1},E_{2}\}}$\tabularnewline
\cline{2-21} 
 & \multirow{6}{*}{$c_{1}c_{2}>1$} & \multicolumn{15}{c|}{$\mathbf{q<p<a<b}$ } & $\emptyset$ & $\mathbf{E_{1}}$(Fig. \ref{fig:partable_mutstrong}c) & $\emptyset$ & $\mathbf{\{E_{1},E_{2}\}}$(Fig. \ref{fig:partable_mutstrong}b)\tabularnewline
 &  & \multicolumn{15}{c|}{$\mathbf{a<b<q<p}$ } & $\emptyset$ & $\emptyset$ & $\mathbf{E_{2}}$(Fig. \ref{fig:partable_mutstrong}g) & $\mathbf{\{E_{1},E_{2}\}}$(Fig. \ref{fig:partable_mutstrong}d)\tabularnewline
 &  & \multicolumn{15}{c|}{$\mathbf{q<a<b<p}$ } & $\emptyset$ & $\emptyset$ & $\emptyset$ & $\mathbf{\{E_{1},E_{2}\}}$\tabularnewline
 &  & \multicolumn{15}{c|}{$\mathbf{a<q<p<b}$ } & $\emptyset$ & $\emptyset$ & $\emptyset$ & $\mathbf{\{E_{1},E_{2}\}}$\tabularnewline
 &  & \multicolumn{15}{c|}{$\mathbf{q<a<p<b}$ } & $\emptyset$ & $\emptyset$ & $\emptyset$ & $\mathbf{\{E_{1},E_{2}\}}$\tabularnewline
 &  & \multicolumn{15}{c|}{$\mathbf{a<q<b<p}$ } & $\emptyset$ & $\emptyset$ & $\emptyset$ & $\mathbf{\{E_{1},E_{2}\}}$ (Fig. \ref{fig:partable_mutstrong}a)\tabularnewline
\hline 
\end{tabular}
\par\end{centering}
\caption{\label{tab:stable_equilibria}Classification of all possible stable equilibria of
model (\ref{eq:plant_odes}) with adaptive animal behavior for all generic parameter
cases.}
\end{sidewaystable}
\par\end{center}

\section{Gradual change in preference\label{sec:appgradual}}

Preference modeled by equation (\ref{eq:pref_step}) in the main text assumes ideal
animals that are omniscient and perfect optimizers that switch instantaneously on
the plant that is more profitable. Now let us consider a more realistic animal that
adjusts its plant preferences more gradually with changes in plant densities. This
can be modeled by the Hill function

\begin{equation}
u_{1}(P_{1},P_{2})=\frac{(e_{1}P_{1})^{z}}{(e_{1}P_{1})^{z}+(e_{2}P_{2})^{z}}\label{eq:pref_hill}
\end{equation}

\noindent where the exponent $z>0$ controls the steepness of preference transitions.
As $z$ converges to infinity, graphs of the Hill functions converge to the graph
of the step-like preference (\ref{eq:pref_step}) in the main text. When we substitute
this gradual switching function in the Lotka\textendash Volterra equations (\ref{eq:plant_odes})
of the main text, piece-wise isoclines change into smooth curves. As the steepness
exponent $z$ increases and switching becomes more step-like, these isoclines converge
to generalized isoclines from the main text.

\begin{figure}
\begin{centering}
\includegraphics[height=0.9\textheight]{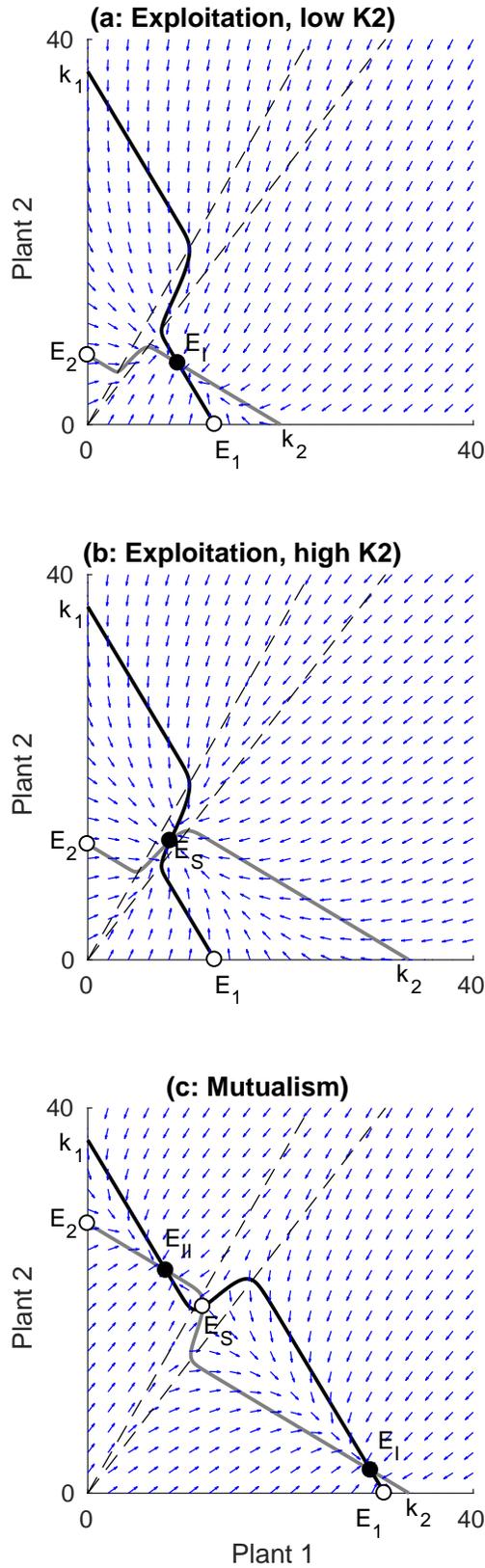} 
\par\end{centering}
\caption{\label{fig:HIclines}Plant isoclines (plant 1: black, plant 2: gray) and population
dynamics under weak competition ($c_{1}c_{2}<1)$ and preferences given by the Hill
function with $z=20$ (dashed lines correspond to contour lines for which $u_{1}=0.05$
and $u_{1}=0.95$). Panels and parameters correspond to those of Figure 2 in the
main text.}
\end{figure}

We observe (cf.~Figure \ref{fig:HIclines} here vs.~Figure \ref{fig:GIclines}
in the main text) that for sufficiently large values of the Hill exponent the dynamics
of model (\ref{eq:plant_odes}) in the main text with step-like preferences are well
approximated by plant population dynamics where animal preferences for plants are
gradual and described by (\ref{eq:pref_hill}). In Figure \ref{fig:HIclines} that
matches Figure \ref{fig:GIclines} of the main text we show a cone of intermediate
plant 1 preferences (area between 5\% and 95\% preference contour lines). Increasing
the Hill exponent $(z)$ towards infinity collapses the cone into the switching line
(equation (\ref{eq:A-switch}) in the main text) and in panel b the intersection
of isoclines converges to $\mathbf{E_{S}}$ given in equation (\ref{eq:E.S}) in
the main text. Isoclines in the cone converge to the segments of generalized isoclines
that are in the switching line.

\end{document}